\documentclass[conference, 10pt]{IEEEtran}
\makeatletter
\def\ps@headings{%
\def\@oddhead{\mbox{}\scriptsize\rightmark \hfil \thepage}%
\def\@evenhead{\scriptsize\thepage \hfil \leftmark\mbox{}}%
\def\@oddfoot{}%
\def\@evenfoot{}}
\makeatother
\usepackage{cite}
\usepackage{amsmath,amssymb,amsfonts}
\usepackage{algorithmic}
\usepackage{textcomp}
\usepackage{xcolor}
\usepackage{tikz}
\usetikzlibrary{arrows.meta, positioning, shapes.geometric}
\ifCLASSOPTIONcompsoc
  \usepackage[caption=false,font=normalsize,labelfont=sf,textfont=sf]{subfig}
\else
  \usepackage[caption=false,font=footnotesize]{subfig}
\fi

\def\BibTeX{{\rm B\kern-.05em{\sc i\kern-.025em b}\kern-.08em
    T\kern-.1667em\lower.7ex\hbox{E}\kern-.125emX}}

\newcommand{\z}{$0^{\text{th}}$ order features}
\newcommand{\one}{$1^{\text{st}}$ order features}

\newcommand{\static}{\textit{Static} scenario}
\newcommand{\mob}{\textit{Mobile} scenario}

\newcommand{\comment}[1]{ }

\newcommand{\myitemizebegin}{\begin{list}{$\bullet$}
{
 \setlength{\leftmargin}{0.4cm}
 \setlength{\parsep}{0.0cm}
 \setlength{\itemsep}{0.05cm}
 \setlength{\topsep}{0.0cm}
}}
\newcommand{\myitemizeend}{\end{list}} 
\begin{document}



\title{Secret-Key Agreement Through Hidden Markov Modeling of Wavelet Scattering Embeddings}

\author{
    \IEEEauthorblockN{Nora Basha\IEEEauthorrefmark{1}, Bechir Hamdaoui\IEEEauthorrefmark{1}, 
    Attila A. Yavuz\IEEEauthorrefmark{2}, 
    Thang Hoang\IEEEauthorrefmark{3}, and Mehran Mozaffari Kermani\IEEEauthorrefmark{2}}
    \IEEEauthorblockA{\IEEEauthorrefmark{1} \textit{School of EECS, Oregon State University, OR, USA}}
    \IEEEauthorblockA{\IEEEauthorrefmark{2}\textit{CS Department, University of South Florida, FL, USA}}
    \IEEEauthorblockA{\IEEEauthorrefmark{3}\textit{Department of Computer Science, Virginia Tech, VA, USA}}
    \textit{\{bashano,hamdaoui\}@oregonstate.edu;attilaayavuz@usf.edu;thanghoang@vt.edu;mehran2@usf.edu}
    }

\comment{
  \author{
 \IEEEauthorblockN{Nora Basha}
 \IEEEauthorblockA{School of EECS \\ Oregon State University, OR, USA\\
 bashano@oregonstate.edu}
 \and
 \IEEEauthorblockN{Bechir Hamdaoui}
 \IEEEauthorblockA{School of EECS \\ Oregon State University, OR, USA\\
  b.hamdaoui@ieee.org}
 \and
 \IEEEauthorblockN{Attila A. Yavuz}
 \IEEEauthorblockA{CS Department \\ University of South Florida, FL, USA\\
 attilaayavuz@usf.edu}
 \and
 \IEEEauthorblockN{Thang Hoang}
 \IEEEauthorblockA{Department of Computer Science \\ Virginia Tech, VA, USA\\
 thanghoang@vt.edu}
 \and
 \IEEEauthorblockN{Mehran Mozaffari Kermani}
 \IEEEauthorblockA{CS Department \\ University of South Florida, FL, USA\\
 mehran2@usf.edu}
 }
}

\maketitle

\begin{abstract}
Secret-key generation and agreement based on wireless channel reciprocity offers a promising avenue for securing IoT networks. However, existing approaches predominantly rely on the similarity of instantaneous channel measurement samples between communicating devices. This narrow view of reciprocity is often impractical, as it is highly susceptible to noise, asynchronous sampling, channel fading, and other system-level imperfections---all of which significantly impair key generation performance. Furthermore, the quantization step common in traditional schemes introduces irreversible errors, further limiting efficiency. In this work, we propose a novel approach for secret-key generation by using wavelet scattering networks to extract robust and reciprocal CSI features. Dimensionality reduction is applied to uncover hidden cluster structures, which are then used to build hidden Markov models for efficient key agreement. Our approach eliminates the need for quantization and effectively captures channel randomness. It achieves a 5x improvement in key generation rate compared to traditional benchmarks, providing a secure and efficient solution for key generation in resource-constrained IoT environments.
\end{abstract}


\section{Introduction}

The broadcast nature of wireless communication, the rapid proliferation of consumer IoT devices, and their constrained resources collectively expose significant security vulnerabilities. These challenges are further exacerbated by the limited scalability and adaptability of traditional public key infrastructure (PKI) and symmetric key cryptographic systems. As a result, there is a growing need for novel encryption, key management, and authentication frameworks tailored for resource-constrained wireless environments. In practice, some IoT and wireless devices rely on cryptographic materials hard-coded into their software or firmware. However, this approach introduces substantial risks, as modern computing hardware can inadvertently leak information through side channel attacks, making pre-stored secrets vulnerable to extraction~\cite{camuratiScreamingChannelsWhen2018}.

Secret-key generation (SKG) techniques based on wireless channels offer a promising alternative for securing IoT devices by harnessing inherent properties of the wireless medium---such as reciprocity and temporal variation---to derive cryptographic key material \cite{mathur_radio-telepathy_2008}.
In most existing SKG approaches, two parties (commonly referred to as Alice and Bob) exploit the randomness of the wireless channel to sample reciprocal parameters. By applying identical quantization schemes to these measurements, they aim to independently generate matching bit sequences, assuming that channel reciprocity yields highly similar instantaneous observations at both ends~\cite{aldaghri_physical_2020, bottarelliAdaptiveOptimumSecret2021}.
However, this assumption often breaks down in practical settings due to factors such as measurement asynchrony, channel noise and fading, hardware imperfections, and packet loss. These challenges frequently lead to key agreement failures and low key generation rates. Moreover, conventional quantization methods inherently introduce errors, further worsening mismatch issues~\cite{basha_wavelet-based_2025,widrow2008quantization}.

In this work, we propose a novel secret-key generation technique that shifts from relying on instantaneous channel measurements to leveraging statistical channel reciprocity. 
It employs a hidden Markov model (HMM)-aided key generation process, where reciprocal clusters found in the low-dimensional channel feature embeddings serve as the HMM states and are used for key generation and agreement. 
Unlike conventional methods, this technique eliminates the need for quantization, as the keys are generated by concatenating the encoded HMM states, offering a robust and efficient alternative for traditional secret-key generation approaches. 


\subsection{Related Works}
Wireless secret-key generation relies on three key assumptions: channel reciprocity and channel temporal variation between the two nodes generating the keys, and spatial decorrelation, which ensures that any other nodes located more than half a wavelength away experience uncorrelated channel conditions \cite{Goldsmith_2005,zhang_key_2016,mathur_radio-telepathy_2008}. However, the asynchronous channel measurements, noise, and impairments in practical Time Division Duplex (TDD) systems deteriorate channel reciprocity which limit the key generation rate and increase bit mismatch between keys.
Extensive research efforts have focused improving the degraded reciprocity of various channel properties, such as the received signal strength (RSS), and channel state information (CSI)~\cite{jiaoPhysicalLayerKey2018, hua_generalized_2023,luo_channel_2023}. In these techniques, signal processing techniques such as Golay filtering \cite{junejoLoRaLiSKLightweightShared2022a}, and discrete cosine transform \cite{margelis_efficient_2019} are used to improve reciprocity and mitigate the impact of noise.
Deep learning techniques have been also proposed for measurements denoising by learning correlated features between the Alice's and Bob's channel measurements. Denoising auto-encoders \cite{chenPhysicalLayerAuthentication2024,zhouPhysicalLayerSecret2022} have been used to encode the channel measurements at both devices into highly correlated representations to improve the key generation and boost the keys' security. Bidirectional convolution neural network has been proposed to increase the key generation rate by minimizing the mean squared error between two devices' channel measurements \cite{chenPhysicalLayerSecretKey2023}. 
%
%
Another direction to increase the key generation rate is based on inducing reciprocal randomness between the communicating devices as proposed in~\cite{liFastSecureKey2022b} where multiple antennas random scheduling and channel obfuscation is used to generate secure, high rate keys. Adaptively shaping the environment to enhance the key generation performance by optimizing Intelligent reconfigurable surfaces (IRS) \cite{hu_reconfigurable_2023} has also shown improved key generation rates. Inducing reciprocal randomness such as CSI manipulation, and permutation at Alice's side, and inferring this induced randomness on Bob's side has been proposed to mitigate the quantization noise and the degraded reciprocity in \cite{du_secret_2024, du_efficient_2025}. Induced random permutations and CSI sequence editing are added by Alice and inferred by Bob using bi-partite graph matching, and the inferred permutation order and editing patterns are used for high rate key generation. 
In most of the key generation techniques, quantization is a fundamental step as it converts the channel measurements into bits, and hence significant activity has been directed to design efficient quantization schemes that reduce the bit mismatch between the reciprocal keys. Lossy adaptive quantization schemes are proposed in \cite{bottarelliAdaptiveOptimumSecret2021,adilAverageContiguousDuration2021} to discard noisy samples and optimally select the quantization threshold values. However, discarding samples lowers the key generation rate, while fixed-level quantization introduces errors and noise, leading to bit mismatches and limiting overall performance.
The traditional consideration of the individual similarities of instantaneous channel measurements samples at the communicating devices has been central in secret key generation literature, limiting the understanding of channel reciprocity. Our proposed approach differ fundamentally from existing key generation in that it considers a statistical view of channel reciprocity, uncovers and leverages hidden embedded reciprocal features within CSI scattering representations.

\subsection{Contributions}
The contributions of this paper are as follows:
\begin{itemize}
   
     \item We propose leveraging wavelet scattering networks to extract and construct CSI feature representations that are reciprocal and robust to time-warping, noise, and fading.
     
    \item We apply dimensionality reduction to reveal and isolate hidden clusters within the high-dimensional CSI data, enabling adherence to reciprocity and temporal variation.
    
    \item We leverage the extracted highly reciprocal clusters to construct hidden Markov models, enabling secret key generation and agreement that achieves a fivefold increase in key generation rate compared to the state of the art.

\end{itemize}

The paper is structured as follows: Section~\ref{sec:proposed} presents an overview of the proposed technique. Section~\ref{sec:features} presents the proposed wavelet scattering network-based construction of CSI features representations. Section~\ref{sec:tsne} presents the proposed CSI feature embedding technique. Section~\ref{sec:hmm} presents the proposed HMM-aided key generation technique and Section~\ref{sec:results} presents our experiment setting and performance evaluation results. Finally, Section~\ref{conc} concludes the paper. 

\section{Overview of the Proposed SKG Technique}
\label{sec:proposed}
\begin{figure*}
\centering
\includegraphics[width=\textwidth]{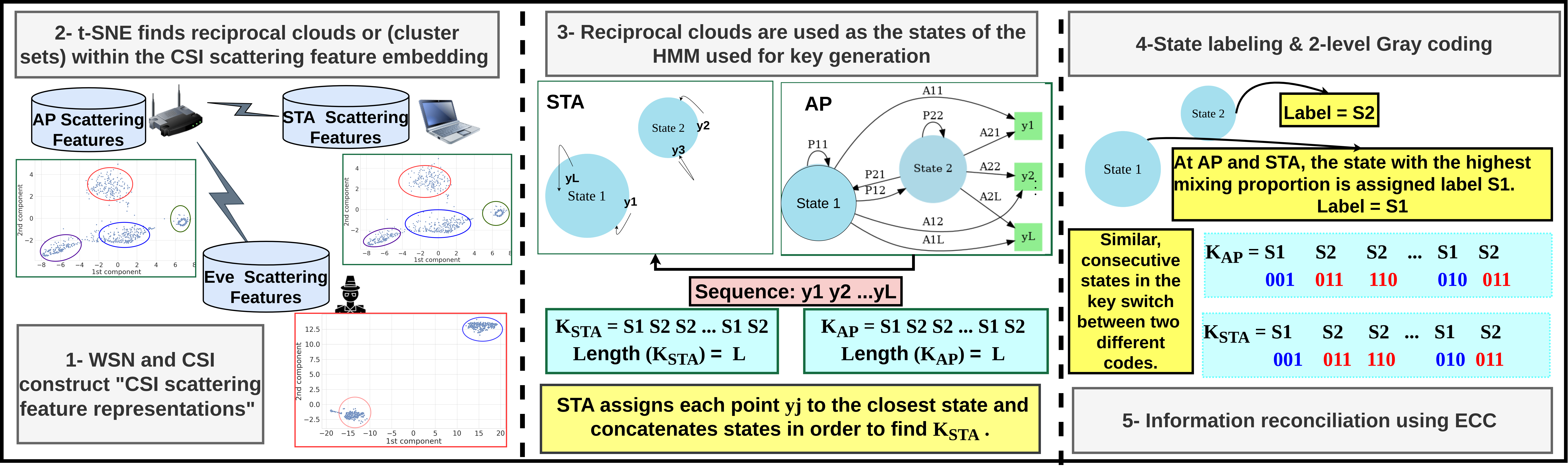}
\caption{Overview of the proposed secret-key generation approach.}
\label{fig:overview}
\end{figure*}

Previously proposed secret-key generation (SKG) approaches focused on designing efficient quantization schemes to convert reciprocal channel measurements, such as RSS or CSI, into bits. This process often involved discarding noisy samples to minimize bit mismatches, which results in the omission of CSI samples. Additionally, quantization typically maps samples of instantaneous channel measurements to a fixed number of levels, introducing quantization errors that contribute to increased bit mismatches between the keys generated at the two end devices. 
In networks of low-cost wireless devices, where noise heavily distorts channel measurements and reciprocity is limited, the reliability of secret key generation further degrades.
To address these challenges, we propose a novel SKG approach that exploits the statistical reciprocity of the wireless channel rather than relying on instantaneous measurements. This method removes the need for quantization, maximizes the use of available channel samples, and significantly enhances key generation rates.


The overview of the proposed SKG scheme is depicted in Fig.~\ref{fig:overview}. The system aims at generating secret keys between two devices, hereafter referred to as Access Point (AP) and Station (STA), using their collected CSI data samples. 
AP and STA each begins (Step $1$ in the figure) by constructing new representations of the collected CSI samples using Wavelet Scattering Networks (WSNs), which are proven to extract time-warping-invariant CSI feature representations that are immune to white noise and amplitude modulation as will be explained in Sec.~\ref{sec:features}. These new representations will be referred to as {\em CSI scattering feature representations}.
%
Next, in Step $2$, t-SNE (t-distributed Stochastic Neighbor Embedding)~\cite{van2008visualizing} is employed to find and extract \textit{reciprocal CSI scattering feature embeddings} from the CSI scattering representations constructed at AP and STA. As shown later, t-SNE visualization reveals that the cluster structures of the feature embeddings extracted by AP and STA are closely aligned and reciprocal, while those extracted by an eavesdropper (Eve), positioned at least half a wavelength away, differ significantly.
Our experiments show that these scattering feature embeddings also evolve over time, reflecting the core principles of secret key generation, namely reciprocity, spatial decorrelation, and temporal variation \cite{zhang_key_2016}. 
A detailed illustration of the extraction of the reciprocal feature embeddings is provided in Section~\ref{sec:tsne}.
 
AP initiates then (Step $3$) the key generation process by constructing a hidden Markov Model (HMM) \cite{murphy2012chapter17} using the extracted reciprocal CSI scattering feature embeddings. The model emits observation sequences, and the AP establishes its key by concatenating the coded states associated with the emitted observation sequence. STA then uses the same sequence and its constructed reciprocal states to deduce the key at AP. 
In Step~$4$, state labeling is performed using our proposed state labeling approach to ensure consistent labeling of states at both the AP and STA. Subsequently, two-level Gray coding is applied to enhance the randomness of the generated keys.
Further details of the proposed HMM-aided key generation technique are provided in Section~\ref{sec:hmm}.
 
\section{CSI Feature Representation Through Wavelet Scattering Networks}
\label{sec:features}
In communication systems where uplink and downlink share the same frequency band, and under rich scattering conditions, CSI samples per subcarrier are expected to exhibit strong correlation (i.e., reciprocity) between the uplink and downlink within the channel's coherence time.
However, factors such as CSI measurement noise, asynchronous sampling, and hardware imperfections introduce unwanted variability that weakens the correlation between uplink and downlink CSI samples, thereby limiting the effectiveness of secret key agreement \cite{bashaEnhancingWirelessSecretKey2024, basha_wavelet-based_2025}. These challenges highlight the need to extract robust channel propagation features that remain consistent across uplink and downlink despite such distortions. In this section, we first identify the key sources of variability that degrade CSI reciprocity. We then present a WSN architecture designed with an appropriate processing order and invariance scale to extract CSI features that are highly correlated and reciprocal across both links. For evaluation, we collect CSI data in a WiFi setup comprising an AP and a STA, both implemented using ESP32 devices. These devices exchange probe packets and independently estimate CSI from the received signals.

\begin{figure} 
    \centerline{
  \subfloat[Raw CSI data containing noise and amplitude modulation distortion \label{subfig:noise}]{%
        \includegraphics[height = 3 cm, width=\columnwidth]{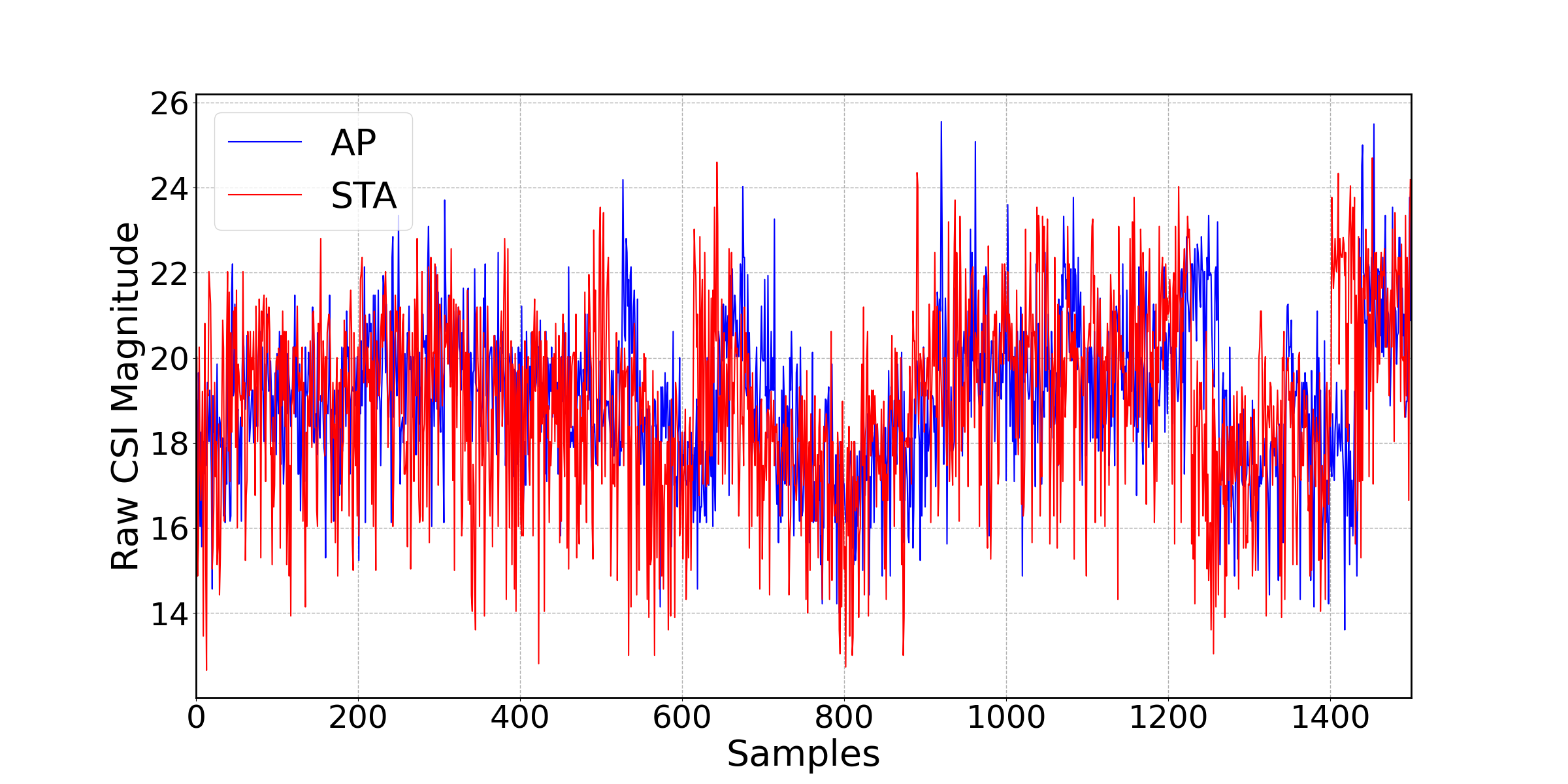}}}

 \centerline{
  \subfloat[Golay-filtered CSI data, showing time-warping as a varying time shift between AP's and STA's CSI data \label{subfig:time_shift}]{%
       \includegraphics[height = 3 cm, width=\columnwidth]{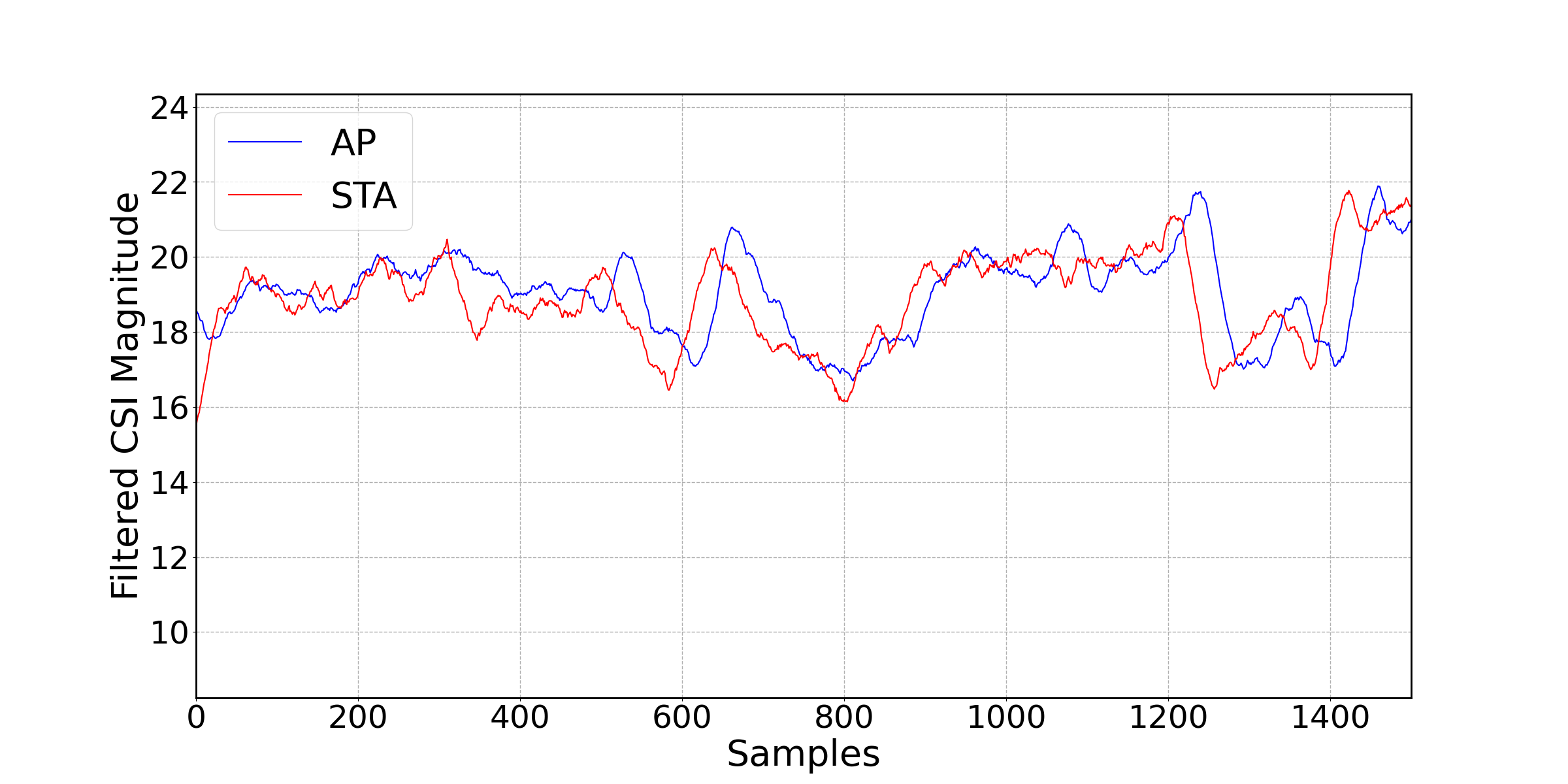}}
       
       }
   \caption{CSI data distortion at AP and STA impacting CSI reciprocity.}
   
  \label{fig:CSI_distortion} 
\end{figure}

\subsection{Time Warping, Noise and Amplitude Modulation}
%
%
Fig.~\ref{fig:CSI_distortion} presents 1,500 raw, unprocessed CSI samples collected at the AP and STA (Fig.\ref{subfig:noise}), along with the same samples after Golay filtering to reduce noise (Fig.~\ref{subfig:time_shift}). 
%
The figure reveals that additive noise, time warping, and amplitude modulation are three key sources of variability and CSI distortion that significantly reduce the correlation and undermine the reciprocity between CSI samples collected by the AP and STA.
Fig.~\ref{subfig:noise} illustrates the impact of additive noise and amplitude modulation, which result in non-reciprocal, fast fluctuations in the CSI magnitudes. Amplitude modulation arises from multipath fading, where signals reflect off multiple obstacles and combine to create rapid amplitude variations. Additional interference and noise sources further amplify this effect.
Fig.~\ref{subfig:time_shift} reveals CSI time warping—visible as a small, varying time shift between the Golay-filtered CSI samples collected at subcarrier 40 by the AP and STA. This shift is caused by the half-duplex nature of WiFi transmissions, which results in the STA's CSI samples being slightly delayed versions of those collected at the AP. Such misalignment leads to increased mismatches in generated key bits when these distorted CSI samples are used in key generation.

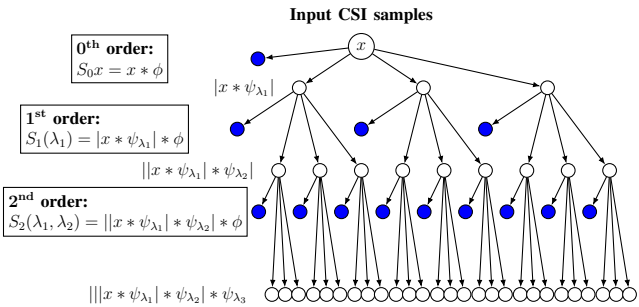
\begin{figure}
\centering
\begin{tikzpicture}[ scale=0.55, transform shape,
  every node/.style={font=\large},
  node distance=1.5pt and 1.5pt,
  level 1/.style={sibling distance=3cm, level distance = 1 cm},
  level 2/.style={sibling distance= 1 cm, level distance = 2 cm,shift={(-1cm,0)}},
  level 3/.style={sibling distance= 0.321cm, level distance = 3 cm, shift={(-0.5cm,0)}},
  edge from parent/.style={draw, -{Latex[length=1mm]}, thin}
]

 \node [circle, draw, fill=white] (x) {$x$}
    child {node [circle, draw, fill=blue, xshift= 2 cm, yshift = 0.7 cm] (n1){}}
    child {node [circle, draw, fill=white, xshift= 0 cm] (na) {}
    child {node [circle, draw, fill=blue, xshift=0 cm, yshift = 1 cm] (n2) {}}
    child {node [circle, draw, fill=white, xshift= 0cm] (nb){}
    child {node [circle, draw, fill=blue, xshift= 0 cm, yshift = 2 cm] (n3){}}
    child {node [circle, draw, fill=white,  xshift= 0 cm] (nc){}}
    child {node [circle, draw, fill=white, xshift= 0 cm] {}}
    child {node [circle, draw, fill=white, xshift= 0 cm]{}}
    }
    child{node [circle, draw, fill=white, xshift= 0 cm] {}
    child {node [circle, draw, fill=blue, xshift= 0 cm, yshift = 2 cm]{}}
    child {node [circle, draw, fill=white, xshift= 0 cm]{}}
    child {node [circle, draw, fill=white, xshift= 0 cm]{}}
    child {node [circle, draw, fill=white, xshift= 0 cm]{}}
    }
    child{node [circle, draw, fill=white, xshift= 0 cm] {}
    child {node [circle, draw, fill=blue, xshift= 0 cm, yshift = 2 cm]{}}
    child {node [circle, draw, fill=white, xshift= 0 cm]{}}
    child {node [circle, draw, fill=white, xshift= 0 cm]{}}
    child {node [circle, draw, fill=white, xshift= 0 cm]{}}
    }
    }
    child{node[circle, draw, fill=white, xshift= 0 cm]{}
    child {node [circle, draw, fill=blue, xshift= 0 cm, yshift = 1 cm]{}}
    child {node [circle, draw, fill=white, xshift= 0 cm] {}
    child {node [circle, draw, fill=blue, xshift= 0 cm, yshift = 2 cm]{}}
    child {node [circle, draw, fill=white, xshift= 0 cm]{}}
    child {node [circle, draw, fill=white, xshift= 0 cm]{}}
    child {node [circle, draw, fill=white, xshift= 0 cm]{}}
    }
    child {node [circle, draw, fill=white, xshift= 0 cm] {}
    child {node [circle, draw, fill=blue, xshift= 0 cm, yshift = 2 cm]{}}
    child {node[circle, draw, fill=white, xshift= 0 cm]{}}
    child {node[circle, draw, fill=white, xshift= 0 cm]{}}
    child {node[circle, draw, fill=white, xshift= 0 cm]{}}
    }
    child {node [circle, draw, fill=white, xshift= 0 cm] {}
    child {node [circle, draw, fill=blue, xshift= 0 cm, yshift = 2 cm]{}}
    child {node [circle, draw, fill=white, xshift= 0 cm]{}}
    child {node [circle, draw, fill=white, xshift= 0 cm]{}}
    child {node [circle, draw, fill=white, xshift= 0 cm]{}}
    }
    }
    child {node [circle, draw, fill=white, xshift= 0 cm]{}
    child {node [circle, draw, fill=blue, xshift= 0 cm, yshift = 1 cm] {}}
    child {node [circle, draw, fill=white, xshift= 0 cm] {}
    child {node [circle, draw, fill=blue, xshift= 0 cm, yshift = 2 cm]{}}
    child {node [circle, draw, fill=white, xshift= 0 cm]{}}
    child {node [circle, draw, fill=white, xshift= 0 cm]{}}
    child {node [circle, draw, fill=white, xshift= 0 cm]{}}
    }
    child{node [circle, draw, fill=white, xshift= 0 cm] {}
    child {node [circle, draw, fill=blue, xshift= 0 cm, yshift = 2 cm]{}}
    child {node [circle, draw, fill=white, xshift= 0 cm]{}}
    child {node [circle, draw, fill=white, xshift= 0 cm]{}}
    child {node [circle, draw, fill=white, xshift= 0 cm]{}}
    }
    child{node [circle, draw, fill=white, xshift= 0 cm] {}
    child {node [circle, draw, fill=blue, xshift= 0 cm, yshift = 2 cm]{}}
    child {node [circle, draw, fill=white, xshift= 0 cm]{}}
    child {node [circle, draw, fill=white, xshift= 0 cm]{}}
    child {node [circle, draw, fill=white, xshift= 0 cm]{}}
    }
    }
    ;

\node[draw, rectangle, align=left, left=2cm of n1] (l0) {$\mathbf{0^{th}}$ \textbf{order:} \\ $S_0 x = x \ast \phi$};

\node[draw, rectangle, align=left, left=1cm of n2] (l1) {$\mathbf{1^{st}}$ \textbf{order:}\\ $S_1(\lambda_1) = |x \ast \psi_{\lambda_1}| \ast \phi$};

 \node[draw, rectangle, align=left, left=0.1cm of n3] (l2) {$\mathbf{2^{nd}}$ \textbf{order:} \\ $S_2(\lambda_1, \lambda_2) = ||x \ast \psi_{\lambda_1}| \ast \psi_{\lambda_2}| \ast \phi$};

\node[above = 0.1 cm of x] {\textbf{Input CSI samples}};
\node[left= 0.3cm of na] {$|x \ast \psi_{\lambda_1}|$};
\node[left= 0.3cm  of nb] {$||x \ast \psi_{\lambda_1}| \ast \psi_{\lambda_2}|$};
\node[left= 0.3cm  of nc] {$|||x \ast \psi_{\lambda_1}| \ast \psi_{\lambda_2}|\ast \psi_{\lambda_3}$};

\end{tikzpicture}
\caption{CSI feature representations using WSN: from input to 2nd-order scattering features}
\label{fig: WSN arch}
\end{figure}


%

\begin{figure} 
    \centerline{
  \subfloat[AP-STA wireless link\label{subfig:AP-STA wireless link}]{%
       \includegraphics[width=0.5\columnwidth]{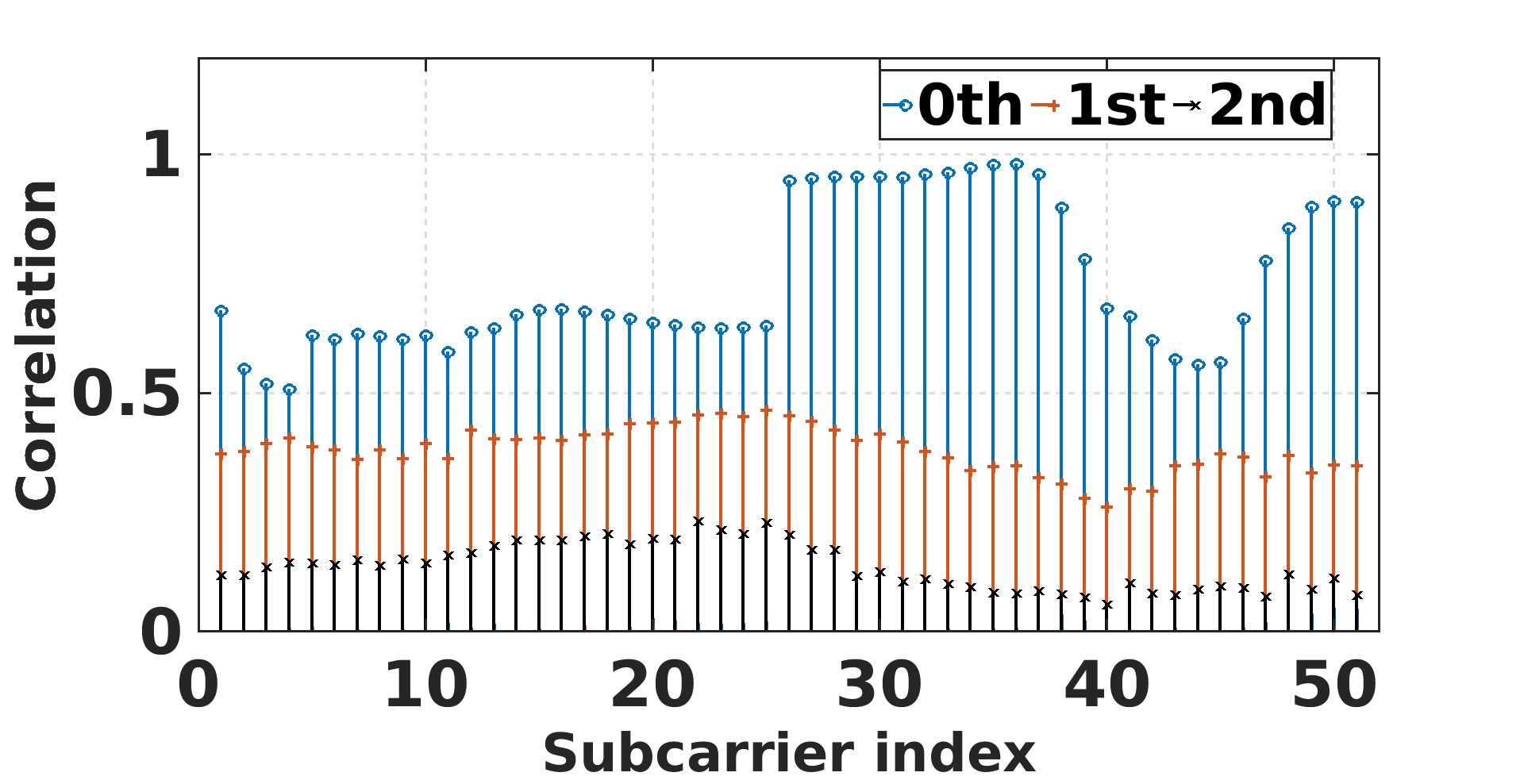}}
  \subfloat[AP-Eve wireless link \label{subfig:AP-Eve wireless link}]{%
        \includegraphics[width=0.5\columnwidth]{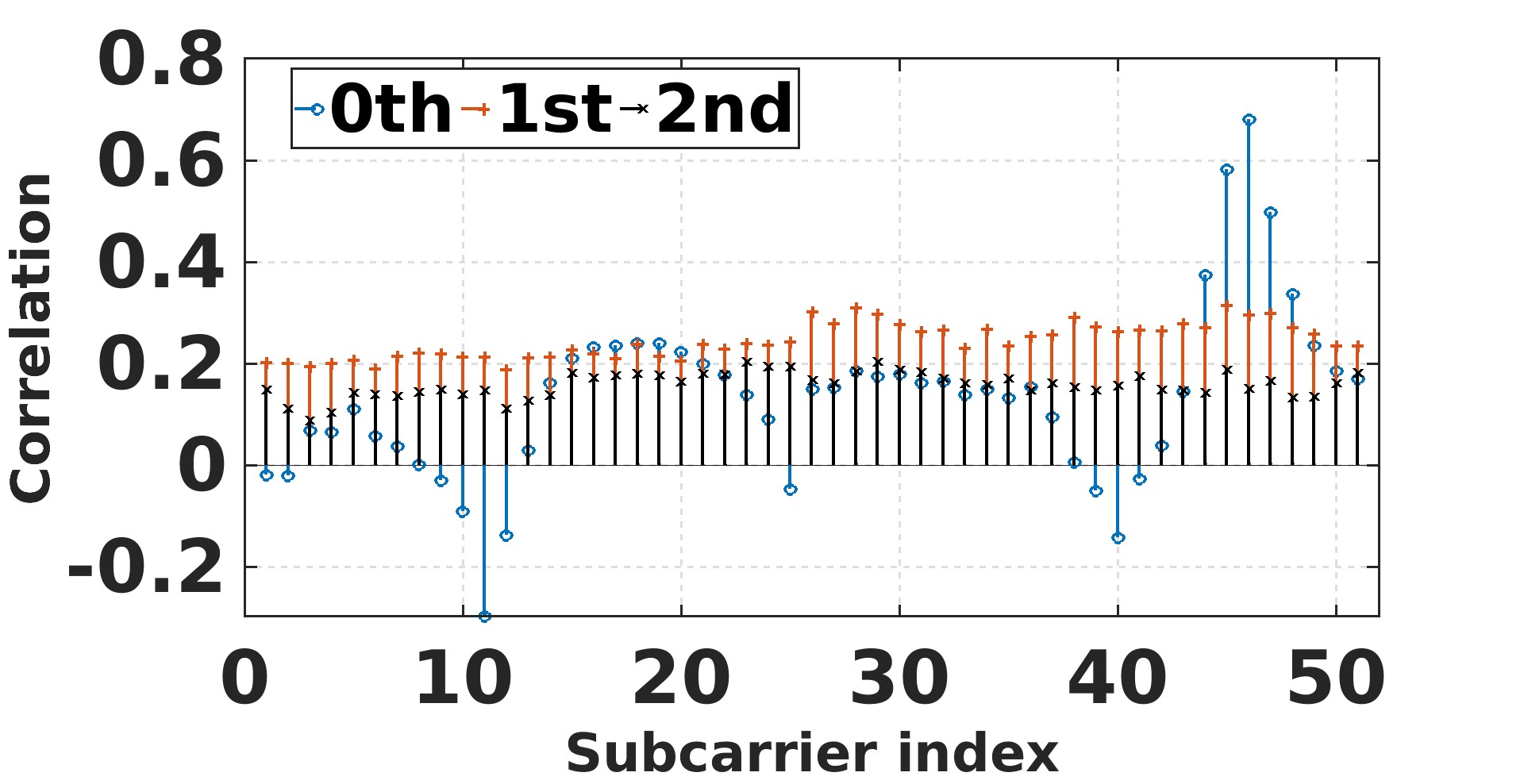}}
}
   \caption{Pearson's correlation of scattering features.}
  \label{fig:WSN_corr} 
\end{figure}

\subsection{CSI Scattering Feature Representations}
In this section, we leverage Wavelet Scattering Networks (WSNs) to address the challenges discussed above. We will begin by introducing WSNs and then explain how they can be exploited to construct CSI feature representations that are resilient to time-warping, noise and amplitude modulation.

In essence, WSNs are a class of deep convolutional neural networks that utilize a fixed cascade of complex wavelet filters and modulus operators, replacing the traditional data-driven linear filters and nonlinear activations \cite{shi_deep_2021}. These predefined filters and non-linearities are designed to yield feature representations that are stable to temporal shifts and deformations in one-dimensional signals, while preserving the intrinsic structure of signal classes \cite{mallat_group_2012, mollai_recursive_2010, singh_deep_2021}.
Because the filters are not learned from data, WSNs are particularly effective in tasks where large datasets are unavailable or when training and testing data come from different domains—as is often the case with CSI data collected over wireless channels \cite{singh_deep_2021}. As a result, choosing appropriate values for the network’s fixed parameters---specifically the network order and the invariance scale \cite{mallat_group_2012}---is crucial to ensuring that the constructed features meaningfully represent the underlying signal characteristics.

WSNs yield identical scattering feature representations for a signal $x(t)$ and its time-shifted version $x(t - \tau)$, provided that the shift $\tau$ is smaller than the support of the network’s invariance scale \cite{anden_deep_2014}. 
This property, known as \textit{time-warping stability}, ensures robustness of the constructed scattering feature representations to small temporal misalignments. Consequently, with an appropriately chosen invariance scale, the constructed features at the AP and STA remain unaffected by the time shifts previously discussed. In our proposed WSN, we set the invariance scale to match the full duration of the CSI signal used for key generation.

Fig.~\ref{fig: WSN arch} illustrates the WSN's produced zeroth, first, and second order scattering features shown as the blue nodes. This WSN comprises two filter banks, $\psi_{\lambda_1}$ and $\psi_{\lambda_2}$ along with the invariance scale $\phi$, and each of filter bank comprises multiple dilated wavelets. Each filter bank, combined with a modulus nonlinearity and followed by averaging over the invariance scale, produces low-variance representations that will be referred to as \textit{CSI scattering feature representations}, with the first filter bank yielding the first-order scattering feature and the second yielding second-order scattering features that capture information lost in the previous layer. The zeroth-order features are obtained by averaging the input CSI signal over the support of $\phi$. While each layer loses some signal details, higher-order features help recover critical structural information, making the representation robust and informative for our task.


\subsection{Reciprocity of CSI Scattering Features}
In this section, we measure and use the correlation as the metric to assess the reciprocity of the CSI scattering feature representations constructed by the AP and the STA.
Fig.~\ref{fig:WSN_corr} presents the correlation of $0^{\text{th}}$, $1^{\text{st}}$, and $2^{\text{nd}}$ order scattering features constructed from CSI data collected over different OFDM subcarriers and over time between AP and STA (Fig.~\ref{subfig:AP-STA wireless link}) and between AP and Eve (eavesdropper) (Fig.~\ref{subfig:AP-Eve wireless link}).

Fig.~\ref{subfig:AP-STA wireless link} demonstrates that for the AP-STA link, $0^{\text{th}}$ order features have the highest correlation between AP and STA over all subcarriers (approximately $0.7$ on average), followed by $1^{\text{st}}$ order features, and then $2^{\text{nd}}$ order features. This observation suggests that lower bit-mismatch rates between AP's and STA's generated keys are expected if $0^{\text{th}}$ and $1^{\text{st}}$ order features are used for key generation. 
The correlation between $2^{\text{nd}}$ order features is consistently less than $0.2$ over all the subcarriers, which confirms that the fast and sudden fluctuation in the CSI signals due to noise and fading amplitude modulation yield not reciprocated features between AP and STA. This also suggests that including $2^{\text{nd}}$ order features in the key generation process leads to increased bit mismatches. Therefore, excluding $2^{\text{nd}}$ order features can enhance key agreement reliability.
Fig.~\ref{subfig:AP-Eve wireless link} shows a remarkably reduced correlation of about $0.2$ between Eve's and AP's CSI scattering features compared to the correlation between AP's and the legitimate STA's constructed features. 
This observation ensures that AP-STA keys are secure against Eve's trials to deduce AP-STA keys by observing the channel. 

\subsection{Construction of Feature Representations} 
\label{subsec:FB}
Since \z~and \one~yield the highest correlation and $2^{\text{nd}}$ order features result in low correlations, we propose to use only the \z~and \one~for key generation and agreement. For this, it suffices to use an WSN with a single filter bank and invariance scale $\phi = S$ with $S$ being the duration of the CSI signal. 

To construct their CSI feature representations, the AP and STA exchange probe packets at a specified time and for a fixed duration, allowing them to sample and collect their respective CSI data. Each device then independently processes the collected CSI to generate its corresponding scattering feature representations. Multiple feature representations can be generated by varying either the subcarrier index and/or the probing times. The total number of generated representations is typically guided by the desired key generation rate.

\section{CSI Scattering Feature Embedding}
\label{sec:tsne}
The CSI scattering feature representations at the AP and STA are high-dimensional datasets, making it difficult to identify patterns and uncover data clusters directly. To address this, we apply dimensionality reduction techniques to extract meaningful clusters from these feature representations. Our results reveal that similar cluster structures emerge in the low-dimensional embeddings of the CSI scattering feature representations at both AP and STA. Furthermore, these cluster sets exhibit reciprocity, spatial decorrelation, and temporal variation properties. These identified clusters serve as the hidden states of the Hidden Markov Model (HMM) employed for key generation and agreement, as detailed in Section~\ref{sec:hmm}.

\subsection{Feature Embedding Through Dimensionality Reduction}
We employ t-distributed stochastic neighbor embedding (t-SNE)\cite{JMLR:v9:vandermaaten08a} to uncover hidden structures and clusters within the high-dimensional CSI scattering feature representations by projecting them into a lower-dimensional space. t-SNE preserves the local structure of the data, ensuring that points that are close in the original space remain close in the reduced-dimensional embeddings~\cite{JMLR:v9:vandermaaten08a}. When applied to the CSI scattering feature representations, t-SNE effectively reveals patterns associated with channel fading and groups features with similar values into well-defined clusters. We refer to these clusters in the low-dimensional embeddings as {\em cluster sets}. These cluster sets exhibit reciprocity between the AP and STA, providing a foundation for developing novel key agreement techniques.

\begin{figure} 
    \centerline{
  \subfloat[AP at T1, sub = 40\label{subfig:tsne-t1-ap}]{%
       \includegraphics[height=3cm,width=0.33\columnwidth]{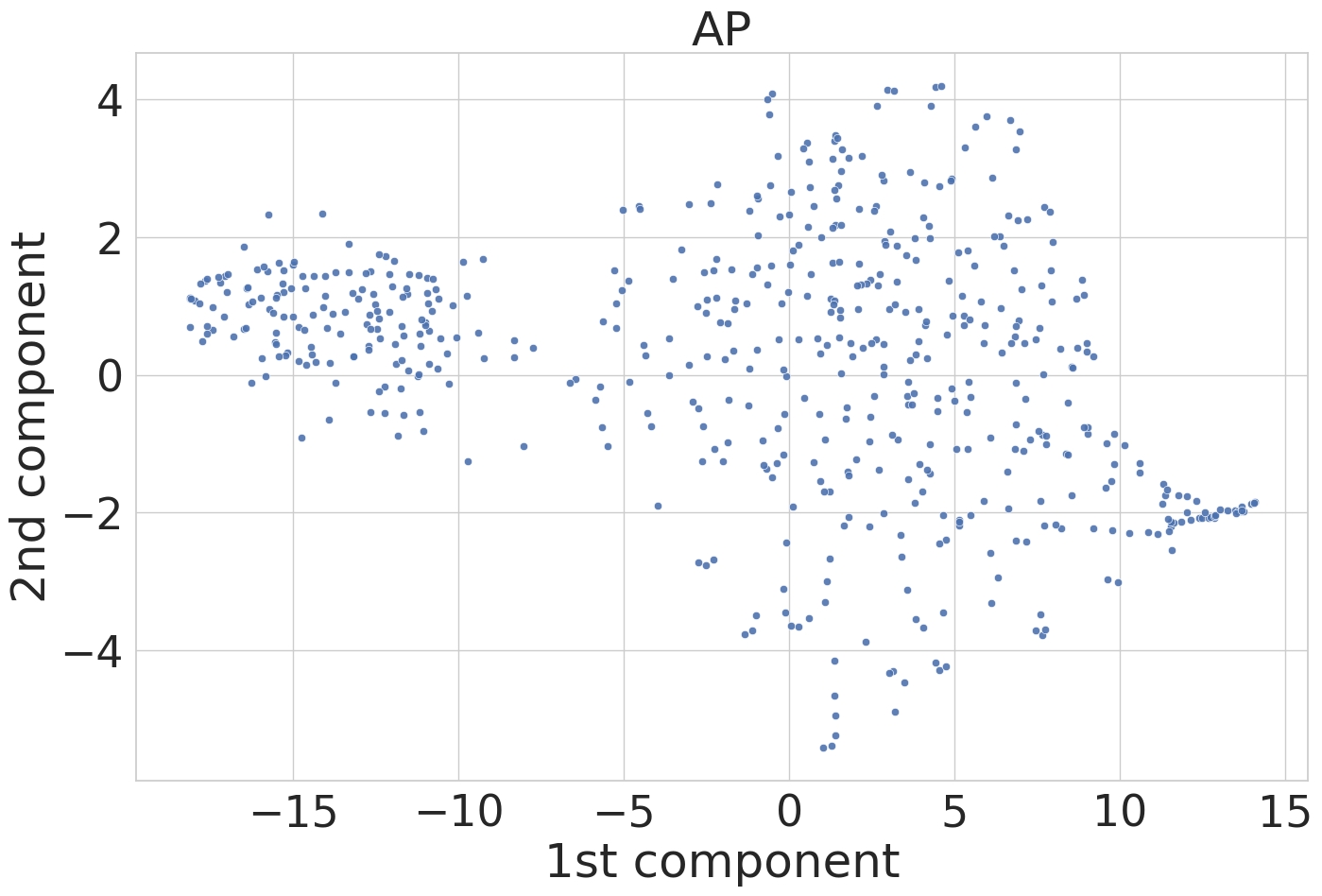}}

  \subfloat[STA at T1, sub = 40 \label{subfig:tsne-t1-sta}]{%
        \includegraphics[height=3cm,width=0.33\columnwidth]{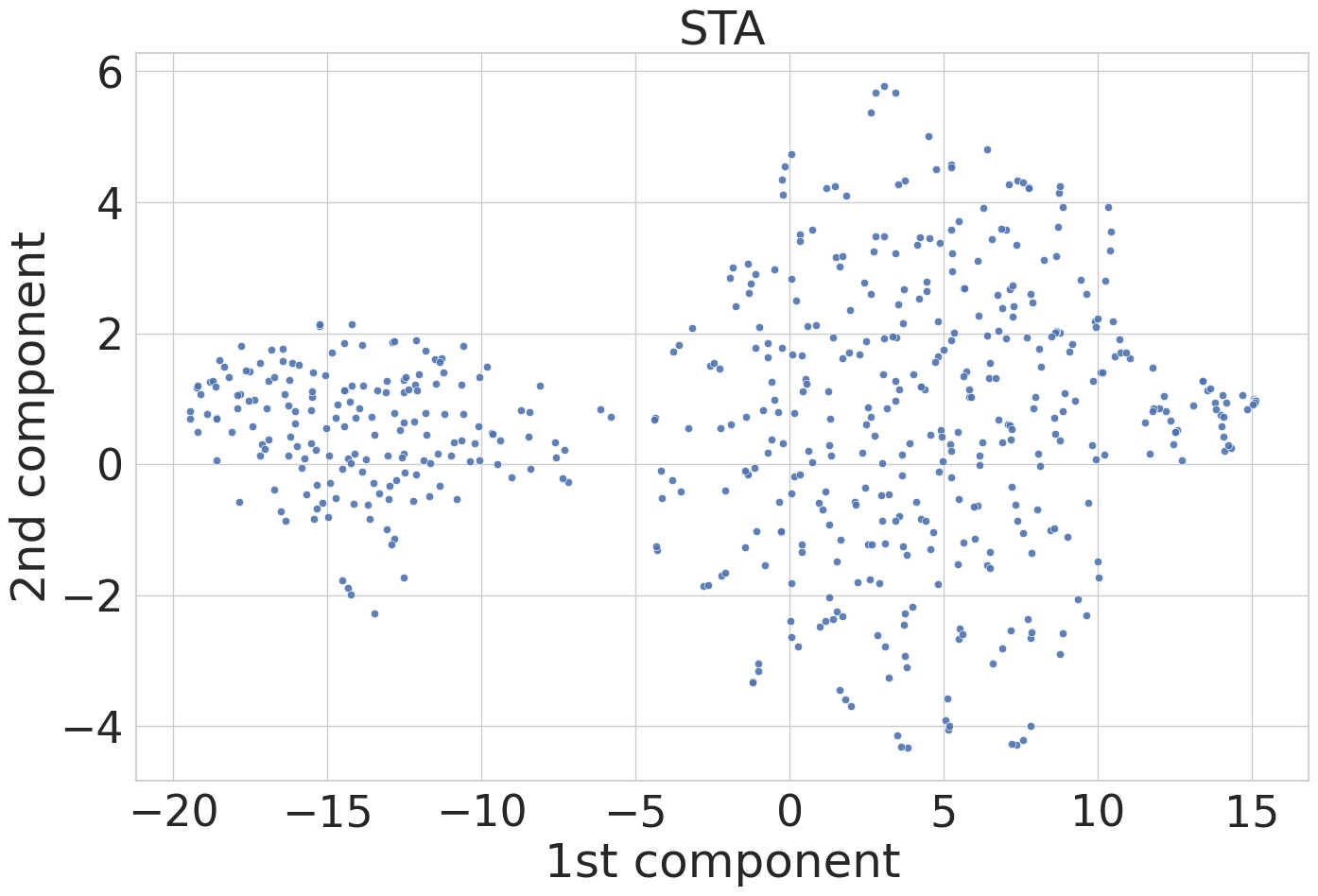}}
    \subfloat[Eve at T1, sub = 40\label{subfig:tsne-t1-eve}]{%
        \includegraphics[height=3cm,width=0.33\columnwidth]{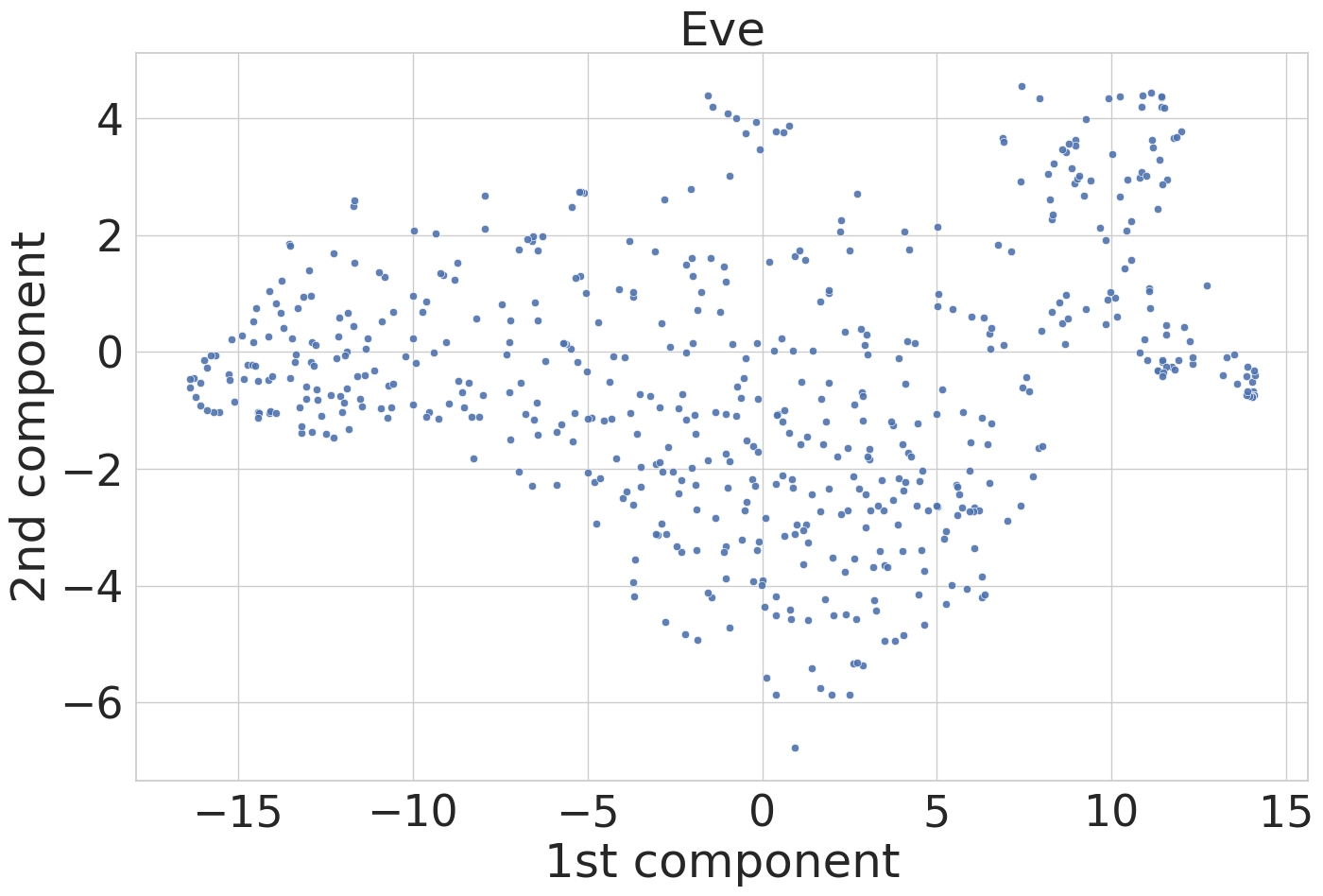}}

  }
  \centerline{
  \subfloat[AP at T2, sub = 40\label{subfig:tsne-t2-ap}]{%
       \includegraphics[height=3cm,width=0.33\columnwidth]{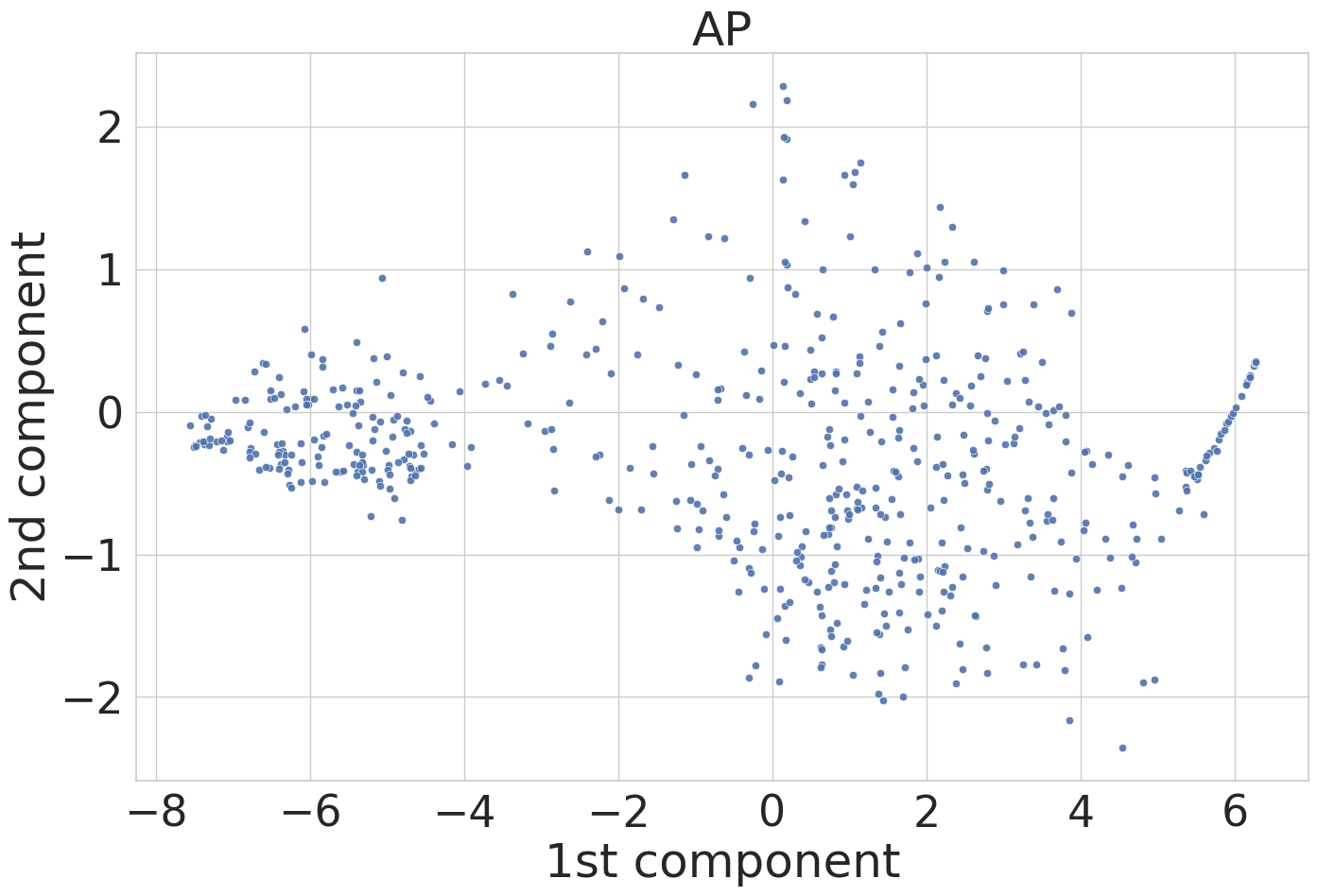}}

  \subfloat[STA at T2, sub = 40\label{subfig:tsne-t2-sta}]{%
        \includegraphics[height=3cm,width=0.33\columnwidth]{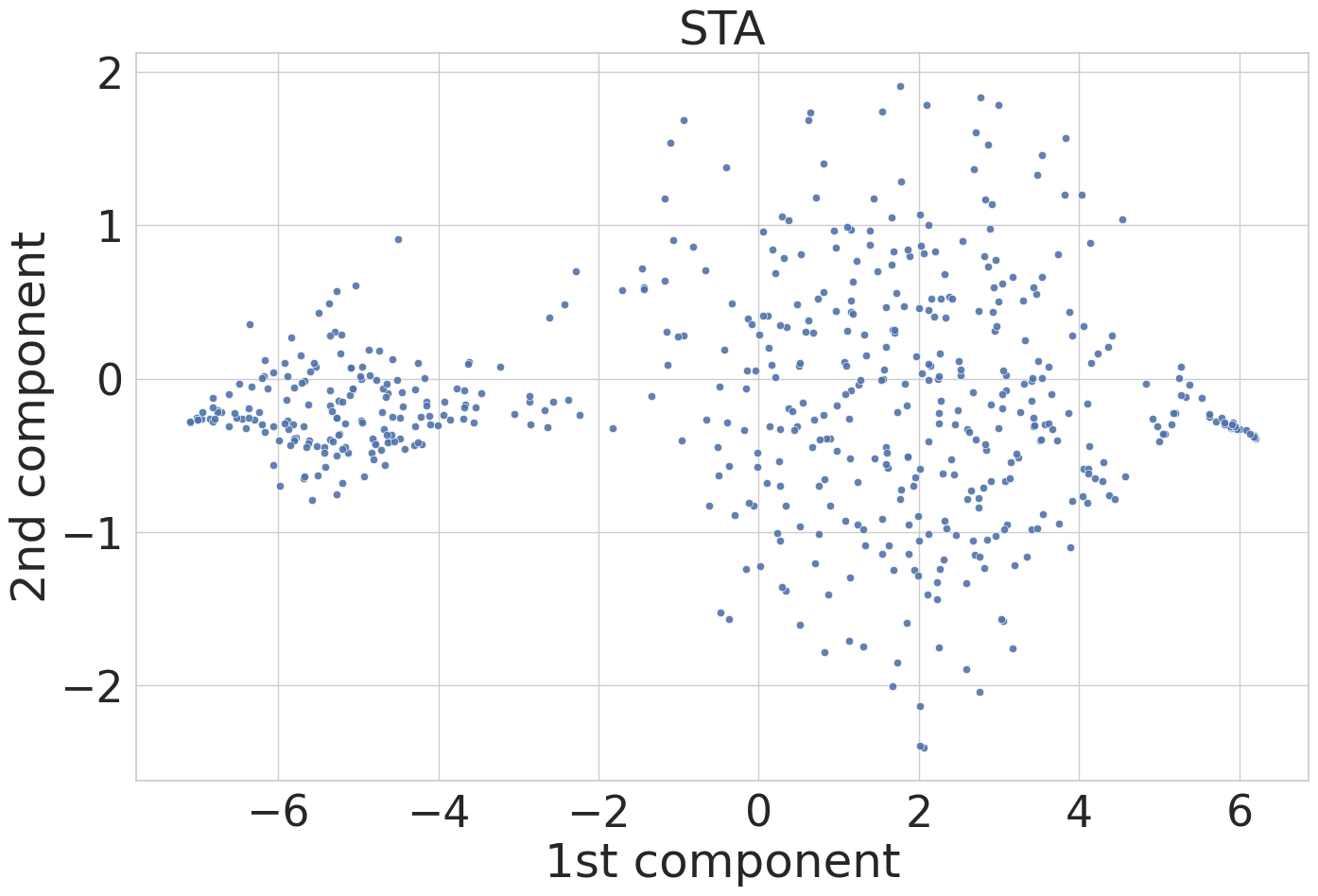}}
    \subfloat[Eve at T2, sub = 40\label{subfig:tsne-t2-eve}]{%
        \includegraphics[height=3cm,width=0.33\columnwidth]{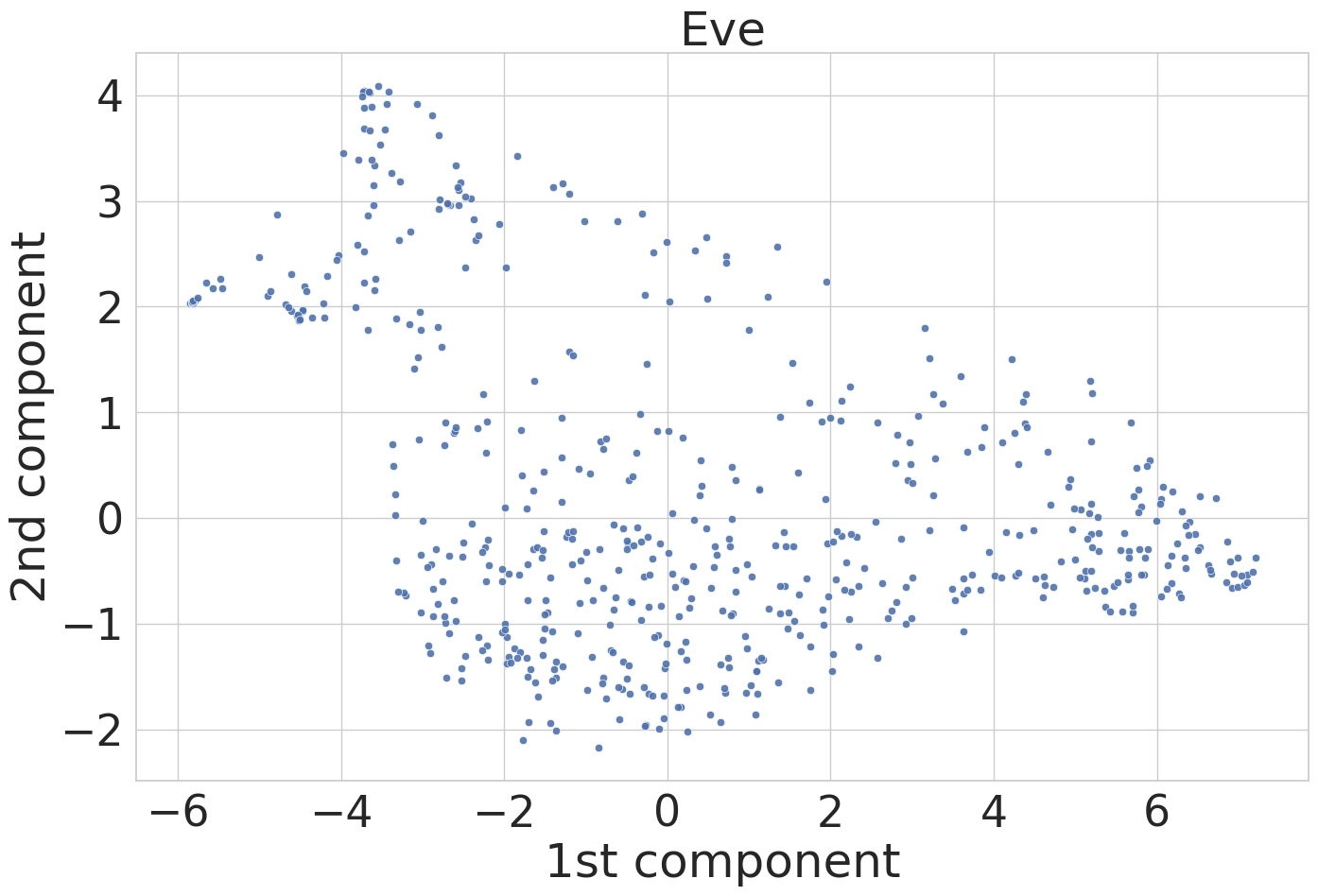}}

  }
  \centerline{
  \subfloat[AP at T2, sub = 50\label{subfig:tsne-sub50-ap}]{%
       \includegraphics[height=3cm,width=0.33\columnwidth]{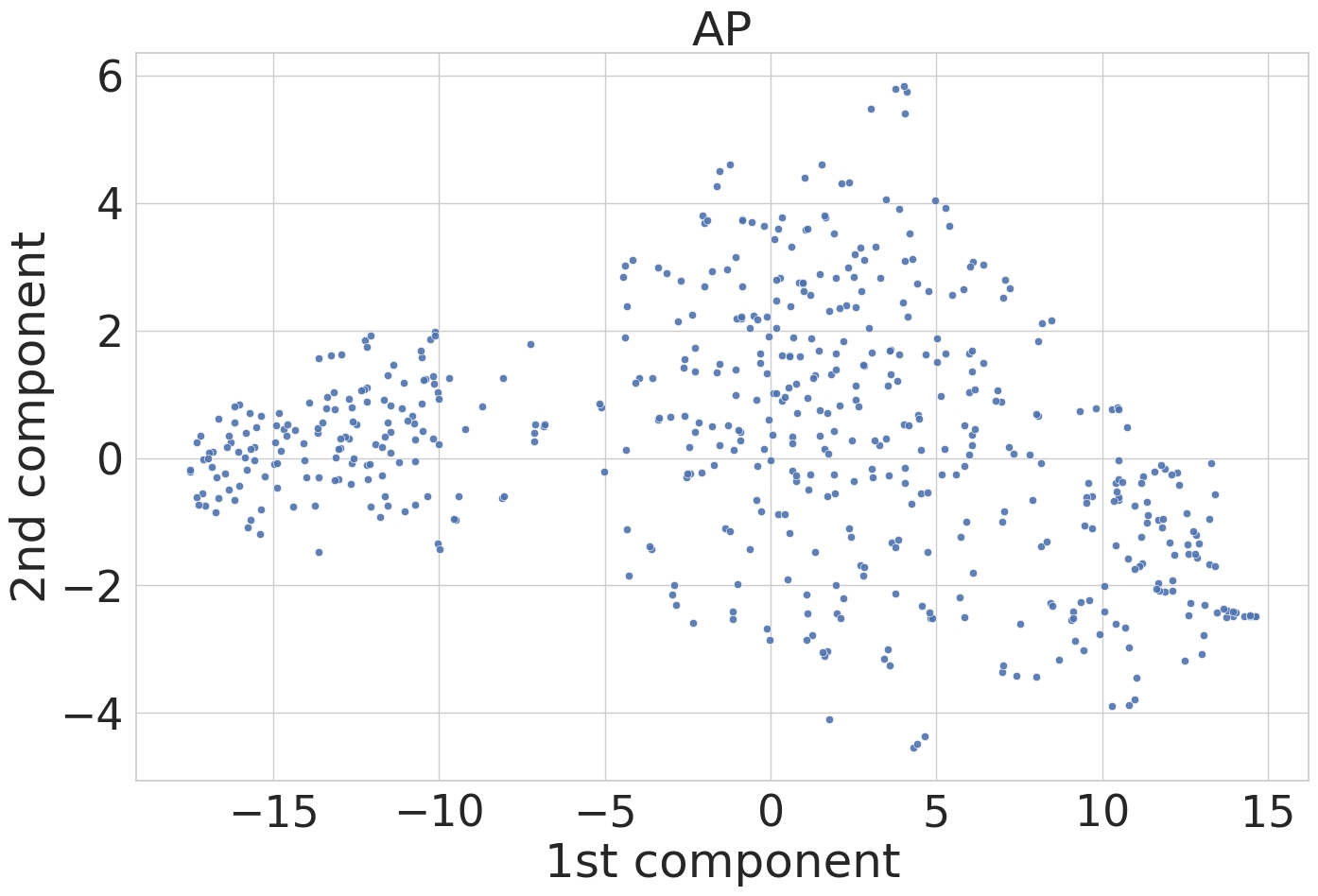}}

  \subfloat[STA at T2, sub = 50\label{subfig:tsne-sub50-sta}]{%
        \includegraphics[height=3cm,width=0.33\columnwidth]{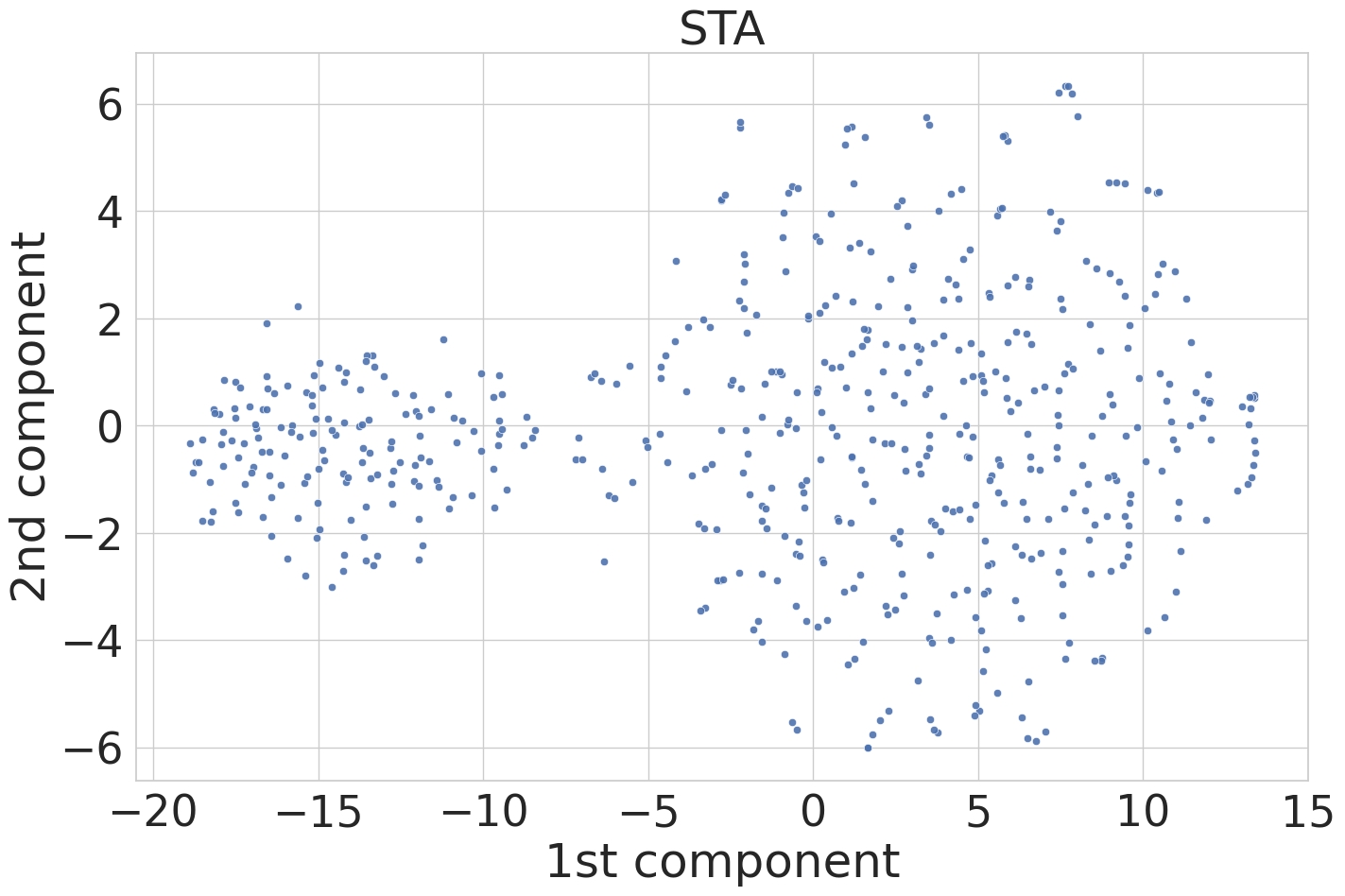}}
    \subfloat[Eve at T2, sub = 50\label{subfig:tsne-sub50-eve}]{%
        \includegraphics[height=3cm,width=0.33\columnwidth]{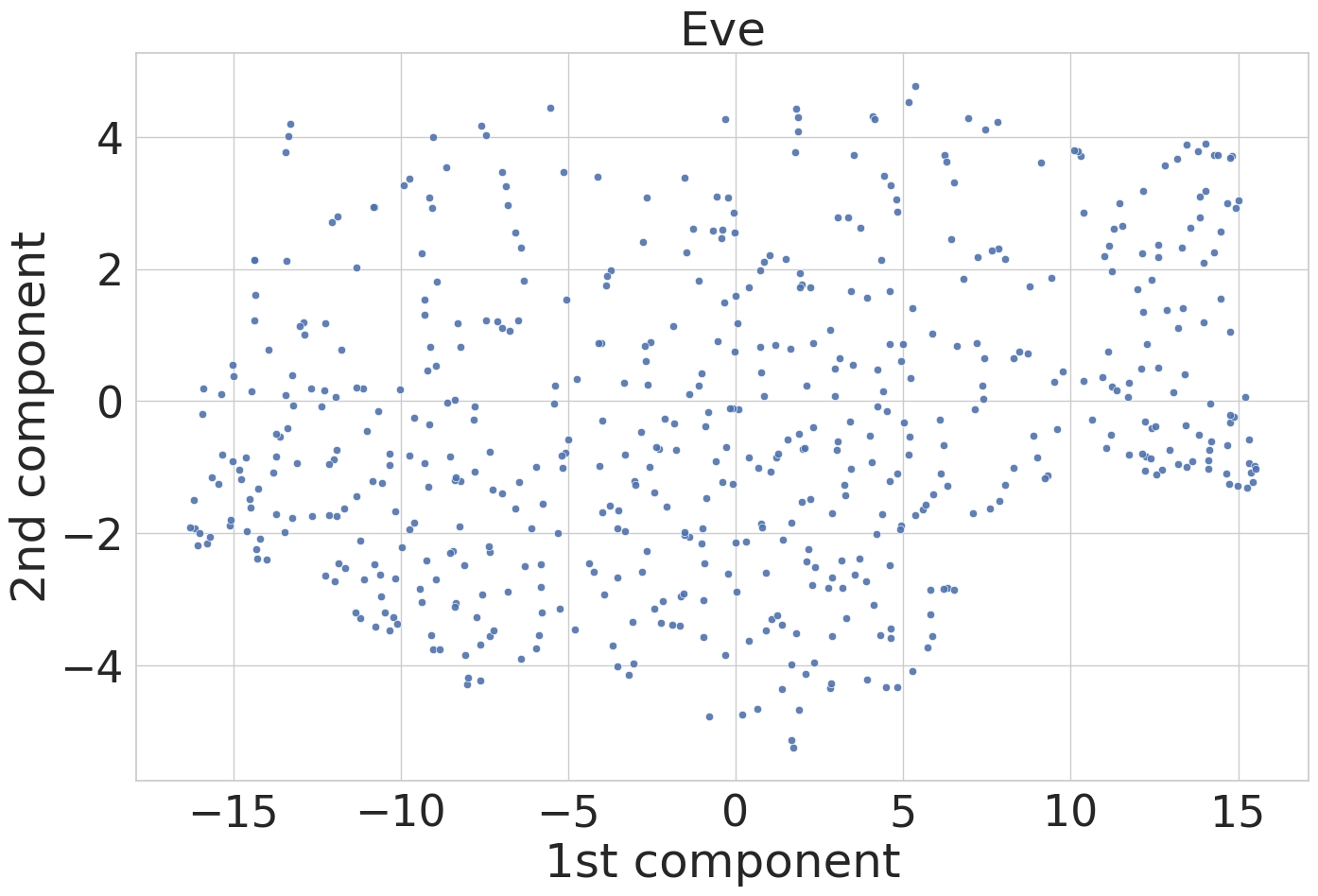}}

  }
  
   \caption{Cluster sets within CSI scattering feature embedding of a {\em single} scattering feature representation. The embeddings are for the scattering representations of $9000$ CSI samples of subcarriers $40$ and $50$ of WiFi at two timestamps T1 and T2.}
  \label{fig:tsne1} 
\end{figure}

Figs.~\ref{subfig:tsne-t1-ap},~\ref{subfig:tsne-t1-sta}, and \ref{subfig:tsne-t1-eve} depict the cluster sets revealed by t-SNE when applied to a \textit{single} CSI scattering feature representation at each of the AP, STA and Eve. Each of these feature representations is of size $24 \times 563$ and is constructed using $9000$ CSI samples collected at time $T_1$ over subcarrier $k = 40$. 
Figs.~\ref{subfig:tsne-t1-ap} and~\ref{subfig:tsne-t1-sta} illustrate closely matching cluster sets at the AP and STA, with both displaying two prominent centroid clusters located approximately at $(-14,1)$ and $(5,1)$. In contrast, Fig.~\ref{subfig:tsne-t1-eve} reveals a distinctly different cluster structure at Eve, despite applying the same construction method to its CSI samples. This divergence remains consistent over time, as shown in Figs.~\ref{subfig:tsne-t2-ap},~\ref{subfig:tsne-t2-sta}, and~\ref{subfig:tsne-t2-eve} for time $T_2$, and across different subcarriers, as seen in Figs.~\ref{subfig:tsne-sub50-ap},~\ref{subfig:tsne-sub50-sta}, and~\ref{subfig:tsne-sub50-eve}  for subcarrier $50$.

The similar cluster structures observed at the AP and STA reflect underlying reciprocal properties that are not evident in the original high-dimensional CSI scattering feature representations and have not been previously utilized for key agreement. Unlike traditional approaches that rely on instantaneous CSI samples—often highly sensitive to noise and responsible for low key generation rates—these reciprocal cluster sets offer greater robustness and reliability.


%

\subsection{Construction of the Cluster Sets}
\label{subsec:CS}
The identified cluster sets can be interpreted as the hidden states of an HMM that emits the CSI scattering feature embeddings. This perspective enables the development of a novel key generation and agreement framework in which both the AP and STA operate and control HMMs to derive matching keys. In this framework, the keys are determined by the sequence of inferred HMM states, as detailed in Section~\ref{sec:hmm}.
Importantly, increasing the number of clusters expands the number of HMM states, thereby enhancing the achievable key generation rate. More clusters can be obtained by generating additional CSI scattering feature representations through measurements over different subcarrier indices and at different probing times.

\begin{figure} 
    \centerline{
  \subfloat[AP\label{subfig:tsne-t12-ap}]{%
       \includegraphics[height=3cm,width=0.33\columnwidth]{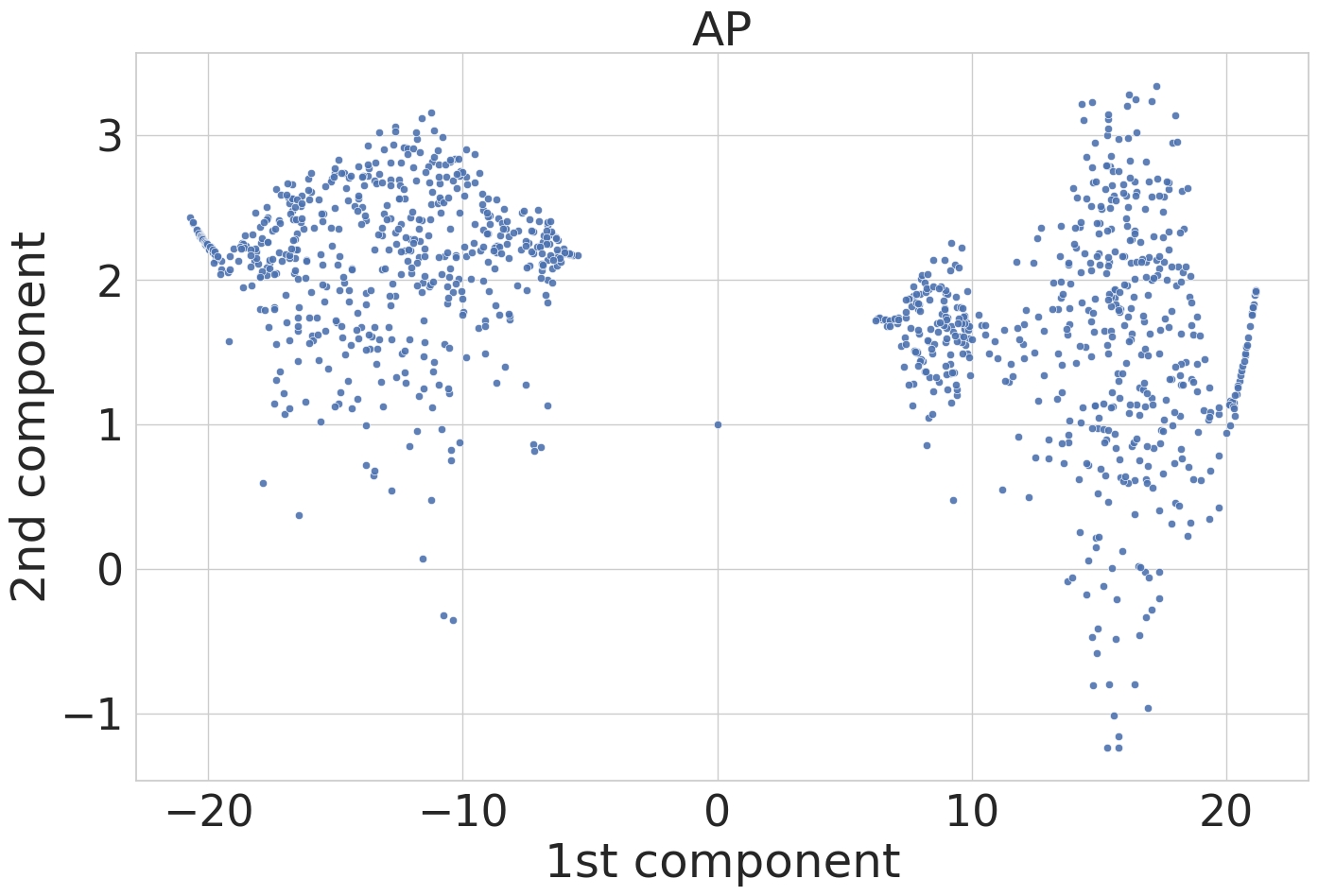}}

  \subfloat[STA \label{subfig:tsne-t12-sta}]{%
        \includegraphics[height=3cm,width=0.33\columnwidth]{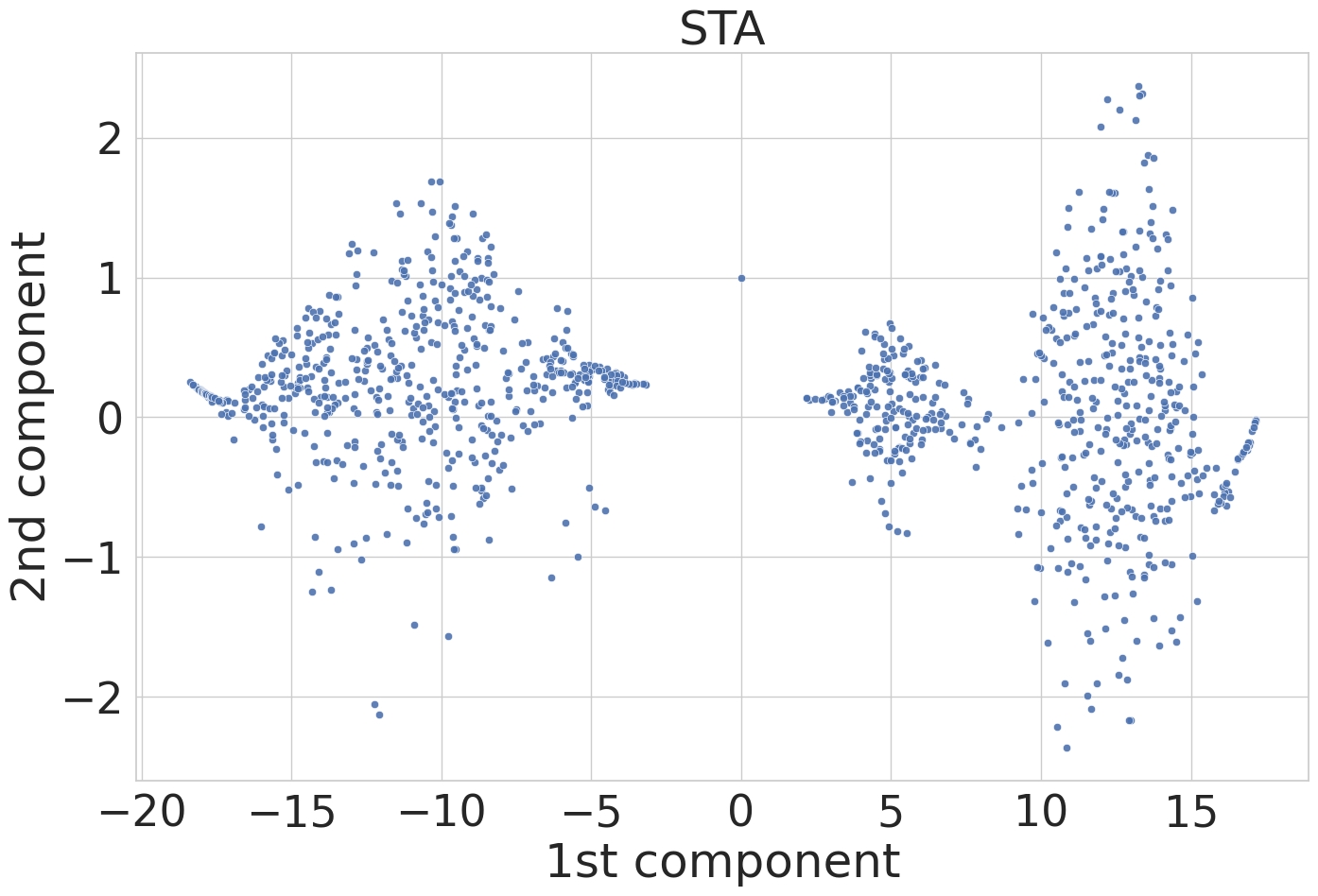}}
    \subfloat[Eve \label{subfig:tsne-t12-eve}]{%
        \includegraphics[height=3cm,width=0.33\columnwidth]{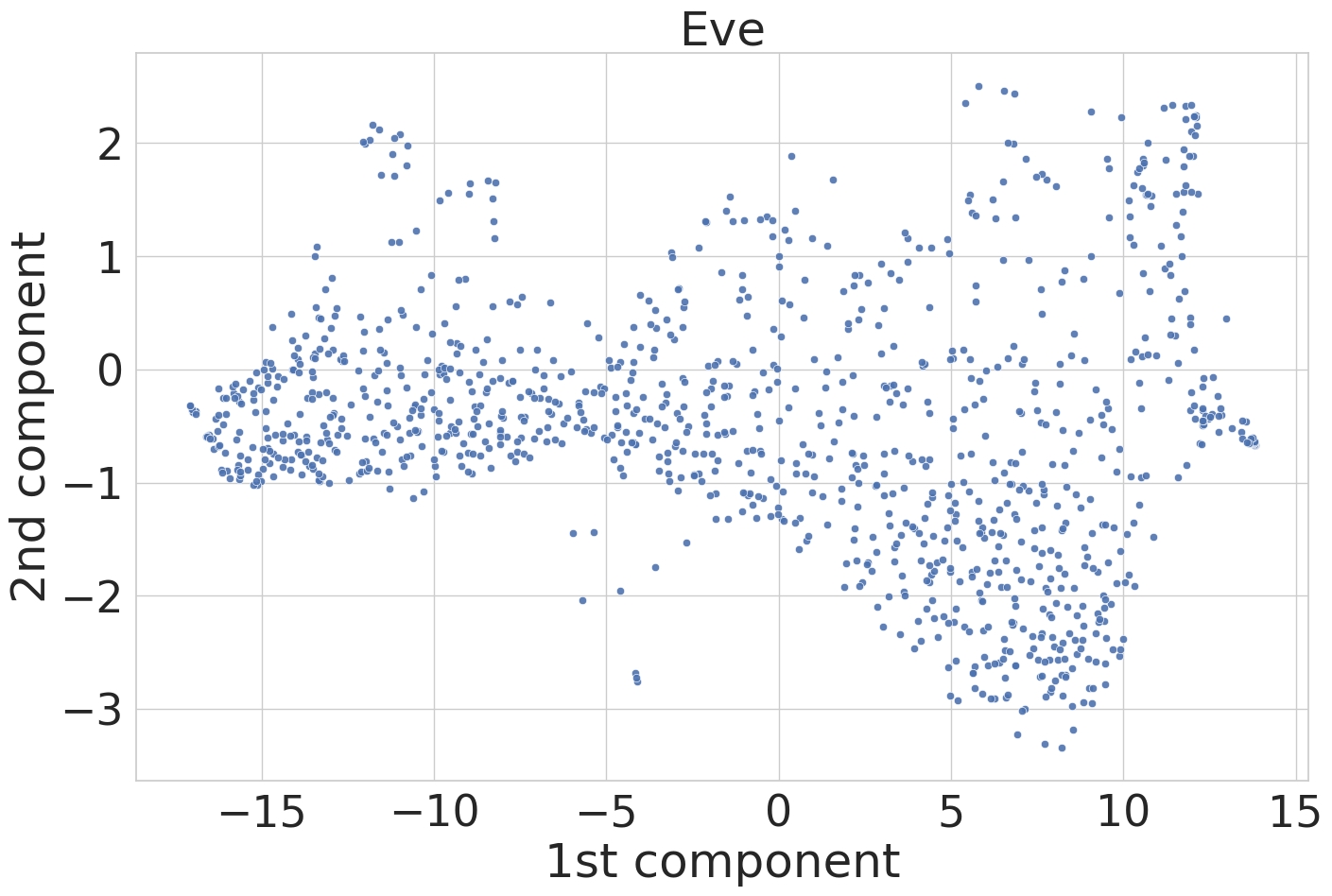}}

  }
  \caption{Clusters extracted from $2$ CSI feature representations constructed from CSI data collected at times $T1$ and $T2$ over subcarrier $40$.}
  \label{fig:tsne2} 
\end{figure}

For example, Fig.~\ref{fig:tsne2}, which shows the cluster sets derived from two CSI scattering feature representations collected at two different times over subcarrier $40$, reveals the emergence of three reciprocal clusters at both the AP and STA. This is in contrast to Fig.~\ref{fig:tsne1}, where only two clusters are observed when using a single feature representation. The figure also highlights that the cluster structure observed at the AP and STA is not mirrored at Eve.

\section{Hidden Markov Modeling and Secret-Key Generation and Agreement}
\label{sec:hmm}
\begin{figure} 

\includegraphics[width = \columnwidth]{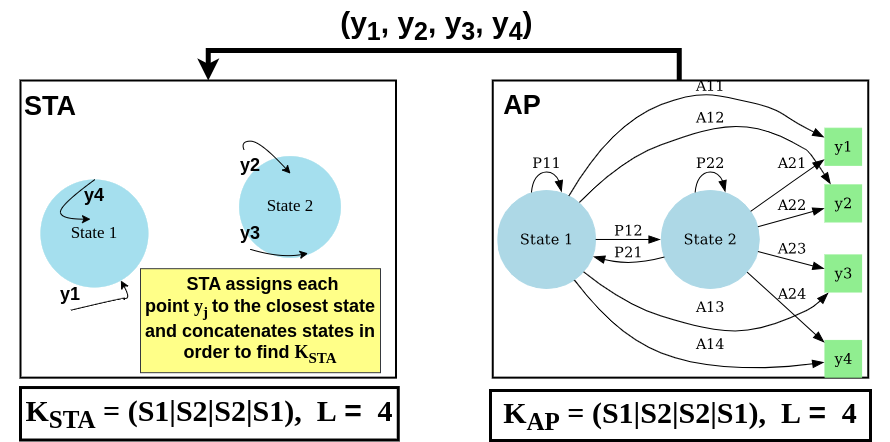}
\caption{An example of two-state hidden Markov model for key generation.}
\label{fig:detail} 
\end{figure}

AP and STA utilize their reciprocal cluster sets to construct individual Hidden Markov Models (HMMs), which serve as the foundation for generating shared secret keys. In this section, we first describe the process by which AP and STA independently build their HMMs using their respective cluster sets, $\mathfrak{C}_{AP}$ and $\mathfrak{C}_{STA}$. We then detail how each device uses its HMM to generate and agree upon secret keys. Next, we outline the clustering method employed to determine the emission distributions associated with each HMM state, based on the CSI scattering feature embeddings. Finally, we introduce a state-labeling technique designed to ensure that corresponding HMM states at AP and STA are assigned consistent labels, enabling reliable key agreement.

\subsection{Hidden Markov Model Construction}
\label{subsec:HMM_construction}
AP constructs its HMM based on its cluster set $\mathfrak{C}_{\text{AP}}$, where each cluster---representing a group of points in the CSI scattering feature embeddings---corresponds to a hidden state in the HMM. 
For illustration, Fig.~\ref{fig:detail} shows an example of an HMM built using two such states. 
The set $\mathfrak{C}_{\text{AP}}$ is used to represent both the set of AP's clusters and its corresponding HMM states.
Once AP constructs its states, it then assigns fixed transition probabilities $Pij$ to transition from State $i$ to State $j$ for all $i,j \in \mathfrak{C}_{\text{AP}}$.
As the HMM transitions from one state to another, it emits observation points according to predetermined state-specific emission probabilities. These probabilities are derived using the Gaussian Mixture Model (GMM) clustering technique~\cite{murphy2012chapter11}, which we describe later in Section~\ref{subsec:GMM}.
STA constructs its own HMM using its cluster set $\mathfrak{C}_{\text{STA}}$, following the same construction methodology as AP.


The key generation process proceeds as follows: AP begins by running its HMM, transitioning between states and emitting a sequence of $L$ observations $O = (y_1, y_2, \dots, y_L)$, where each observation $y_j$ is produced when a particular state is visited. Once the sequence $O$ is obtained, AP generates its key, $K_{\text{AP}}$, by concatenating the labels of the states that emitted the observations, in order. For instance, as illustrated in Fig.~\ref{fig:detail}, if the HMM makes $L=4$ ordered visits---say (State 1, State 2, State 2, State 1), then the generated key is the ordered sequence of labels, $K_{\text{AP}} = (S1|S2|S2|S1)$, where $S_i$ is the label of State $i \in \{1,2\}$. State labeling will be described in Section~\ref{subsec:labeling}.
Finally, AP transmits the sequence of observations $O$ to STA for key generation and agreement.

Upon receiving $O = (y_1, y_2, \dots, y_L)$, STA maps each point $y_i$ to its nearest state and constructs its key, $K_{\text{STA}}$, by concatenating the labels of the inferred states corresponding to the full sequence of mapped points. This ensures that $K_{\text{STA}}$ closely matches $K_{\text{AP}}$, assuming reciprocity holds and the cluster structures at AP and STA are aligned.
Referring back to Fig.~\ref{fig:detail} for illustration, STA maps $y_1, y_2, y_3$ and $y_4$ to its closest states, State 1, State 2, State 2, and State 1, yielding $K_{\text{STA}}=(S1|S2|S2|S1)$ that matches AP's generated key.

\subsection{HMM Emission Probabilities}
\label{subsec:GMM}
Since the HMM states correspond to clusters of points in the CSI scattering feature embeddings, and each emitted point \( y_j \) belongs to one of these clusters, we propose using Gaussian Mixture Model (GMM) clustering~\cite{murphy2012chapter11} to model the emission probability distribution of each cluster. 
Specifically, for a total number $C$ of clusters, the embeddings at AP are modeled as samples from \( C \) Gaussian components, each component $i$ defined by a mean \( \mu^i_{\text{AP}} \), covariance matrix \( \Sigma^i_{\text{AP}} \), and mixing proportion \( p^i_{\text{AP}} \). 
These distributions serve as the emission distributions for the HMM states.
During key generation, AP uses the learned Gaussian mixture distribution to determine the emission probabilities $Aij$ from State $i$ to any point $y_j$. $Aij$ is used to produce the sequence of observations \( O = (y_1, y_2, \dots, y_L ) \), where each \( y_j \) is sampled from the distribution associated with the emitting HMM state. The key \( K_{\text{AP}} \) is then formed by concatenating the labels of the visited states in order.

STA follows the same procedure: it applies GMM clustering to its CSI scattering feature embeddings, resulting in \( C \) Gaussian distributions with parameters \( \mu^i_{\text{STA}} \), \( \Sigma^i_{\text{STA}} \), and \( p^i_{\text{STA}} \). Upon receiving the observation sequence \( O \), STA maps each point \( y_j \) to the nearest Gaussian component in its model and reconstructs its key, \( K_{\text{STA}} \), by concatenating the labels of the corresponding inferred states.

\subsection{State Labeling and Key Generation}
\label{subsec:labeling}
Once AP and STA construct their state sets and their associated Gaussian distributions, they need to agree on the states' labels to ensure key agreement. One straightforward solution is to utilize the means of the Gaussian distributions to produce labels, e.g. label states according to their increasing order of means.
Now since AP and STA have similar cluster structures, they end up having matching labels.
However, we find that the mixing proportions for each of the Gaussian distributions of the GMMs at AP and STA are also reciprocal because of the similar cluster structures at AP and STA. Therefore, we propose to label the states according to the increasing order of the mixing proportions of the associated Gaussian distributions. 
In the proposed labeling technique, at AP, the labeling is performed as follows: Each state is assigned a label value that is proportional to its mixing proportion. 
Since the values of the mixing proportions of states are consistent across both AP and STA, both AP and STA end up with the same labeling. The labeling can, for instance, be done according to the increasing order of mixing proportions.
The same labeling process takes place at STA.  

The proposed labeling technique captures the shape and structure of the clusters at AP and STA (the mixing proportions) and hence is less impacted by noise compared to the states' means. In addition to robustness to channel measurement noise, labeling using mixing proportions cancels the need for the quantization step adopted in traditional secret-key generation techniques and hence is robust to quantization errors.
After labeling, the states at AP and STA are encoded into bits using any encoding scheme such as Gray coding~\cite{bitner_efficient_1976}.

Finally, to obtain the keys as sequences of bits instead of a concatenation of state labels, AP and STA replace each state's label in $K_{AP}$ and $K_{STA}$ with its corresponding Gray code. The devices confirm that the keys are identical through an information reconciliation process using any of the error correction techniques~\cite{zhang_key_2016}.

\section{Performance Results and Analysis}
\label{sec:results}
We assess the performance of the proposed technique by comparing it with two recently proposed key generation schemes: {\em Denoising Autoencoder (AE)} \cite{zhouPhysicalLayerSecret2022} and {\em Bidirectional Convolutional Feature Learning (BCFL)} \cite{chenPhysicalLayerSecretKey2023}.
For key generation performance evaluation, we use the following metrics:
\myitemizebegin
\item \textbf{Bit Error Rate (BER)}, the average number of mismatched bits between the AP's and STA' keys divided by the key length.

\item \textbf{Key Generation Rate (KGR)}, the number of bits generated per probing packet/measurement. 

\item \textbf{Randomness} of the generated keys, which we assess using NIST test suite \cite{8966}.
\myitemizeend

\subsection{Experimental Setup and Scenarios} 
We used a pair of Pycom devices to function as the AP and STA, wirelessly connected via the 802.11n WiFi protocol at 2.427 GHz. CSI samples---comprising the magnitudes of the estimated channel impulse responses across OFDM subcarriers---were collected using the ESP32 CSI Toolkit \cite{Hern2006:Lightweight}. Data collection was conducted simultaneously at the AP and STA in 20-minute blocks, with each block containing 9,000 CSI magnitude samples per subcarrier, and tagged with the timestamp marking the start of the collection.

In our evaluation, we generated keys using CSI samples collected at $2$ different timestamps each over $11$ subcarriers, 
resulting in $121$ different pairs of cluster sets at AP and STA. The reported results show the key performance of the proposed key generation and agreement technique averaged over $5$ CSI samples blocks collected by AP and STA. In this evaluation, we study two scenarios:
\myitemizebegin
\item \static: A line-of-sight (LoS) communication scenario, where the AP and STA are not moving and separated by a distance of 6.5 meters in an indoor environment.
\item \mob: A Non line-of-sight (NLoS) communication scenario, where STA is moving indoors within the AP's connectivity range.
\myitemizeend
%


\subsection{Key Generation Rate \& Bit Error Rate Analysis}
We evaluate the KGR at $5$ different values for the allowed number of mismatched bits between AP's and STA's keys (bits in error) that the information reconciliation step could potentially correct. Throughout the evaluation, we refer to the percentage of mismatched bits between AP's and STA's keys relative to the key length as bit-mismatch percentage, and we consider all the generated key pairs with bits in-error below the studied bit-mismatch percentage values as successful keys.

\begin{figure}
\centering
\includegraphics[width=\columnwidth]{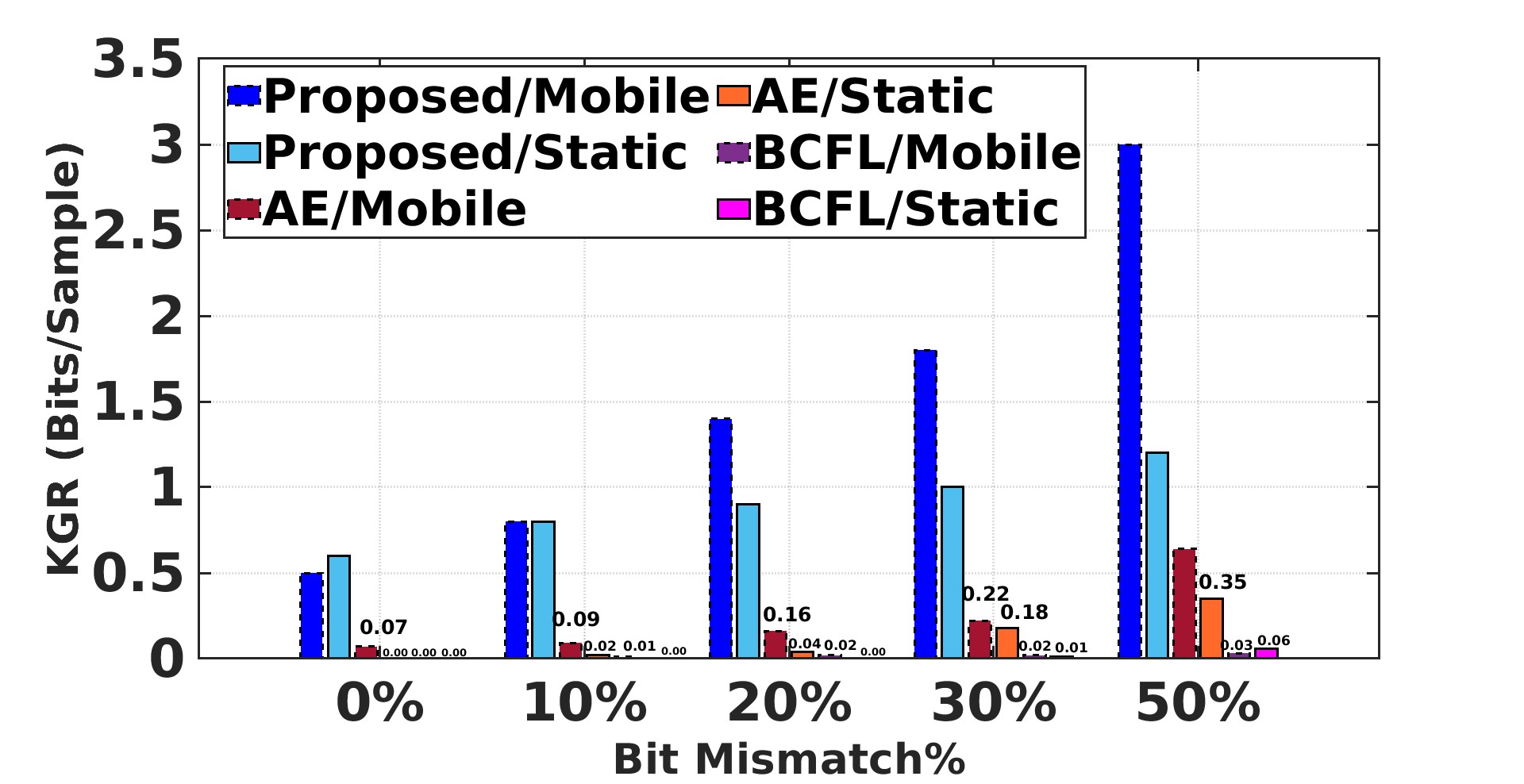}
\caption{Average KGR for 5 different bit mismatch values. $0\%$ corresponds to a perfect key agreement between AP and STA.}
\label{KGR}
\end{figure}

Fig.~\ref{KGR} shows the achieved KGRs in bits per sample at different bit-mismatch values for the studied techniques before information reconciliation.    
The figure illustrates that the proposed HMM-aided key generation has the highest KGR under the static and mobile scenarios compared to the other techniques. For example, when only keys with zero bit mismatch are considered successful, the proposed technique generates up to $0.6$ bits per sample in static environments, while AE and BFCL fail in key generation. 
As the allowed bit mismatch increases, the proposed technique generates up to $3$ bits per sample under \mob~compared to $0.6$ and $0.03$ bits per sample for AE and BCFL, respectively. The KGR of the proposed technique demonstrates that utilizing the reciprocity of the cluster structures within the CSI scattering embeddings, rather than focusing on enhancing the reciprocity of instantaneous samples, shows a significant potential for enabling secret-key generation in low-cost wireless devices, where channel measurements are heavily affected by noise and imperfections. The significantly high KGR of the proposed technique compared to AE and BCFL key generation is also attributed to utilizing different combinations of scattering features representations of different timestamps and subcarriers to generate multiple cluster sets---and more keys---from the same probing packets. 

\begin{figure} 
    \centerline{
  \subfloat[AP\label{subfig:ap_mob}]{%
       \includegraphics[ height=3.3cm,width=0.33\columnwidth]{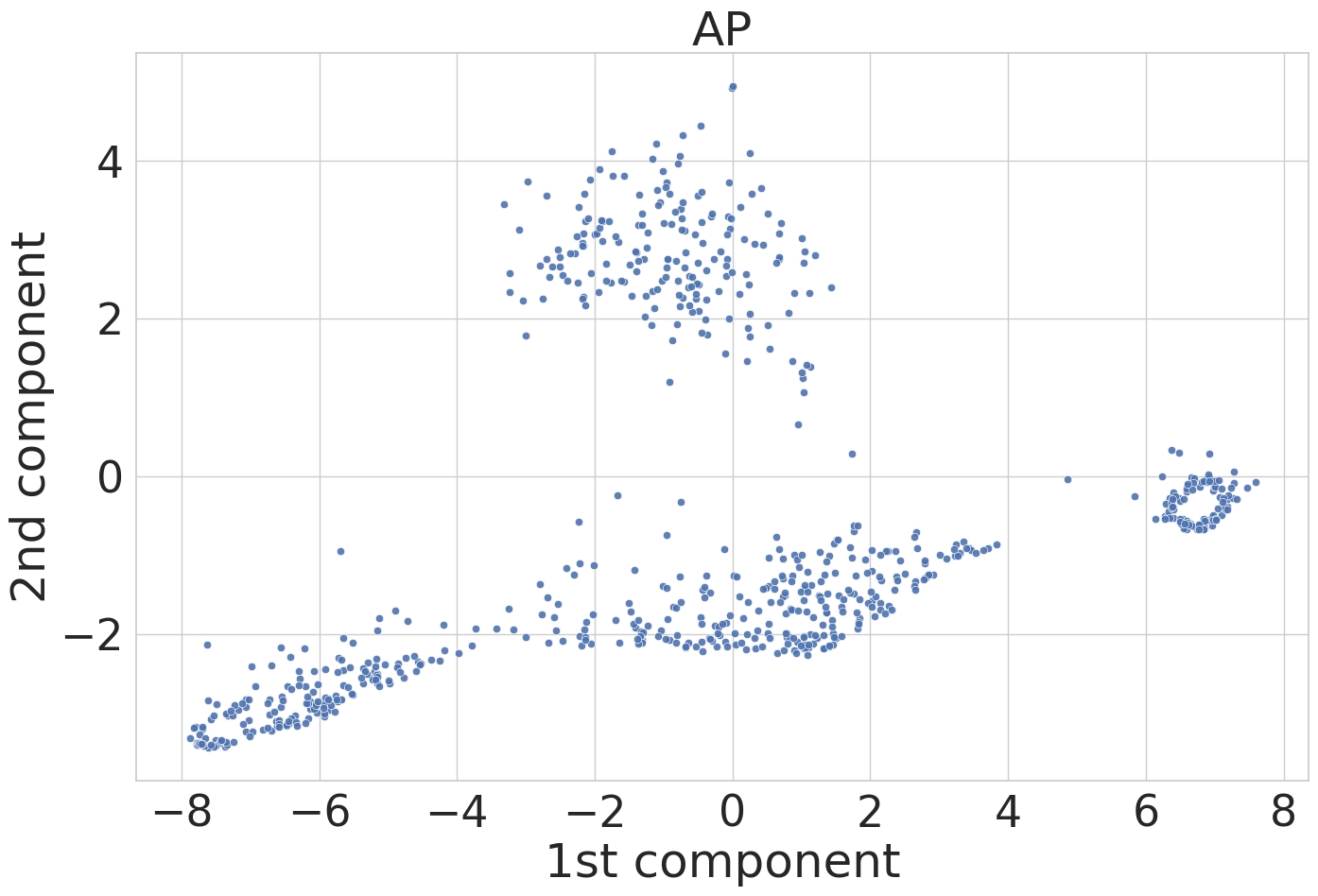}}
  \subfloat[STA \label{subfig:sta_mob}]{%
        \includegraphics[ height=3.3cm,width=0.33\columnwidth]{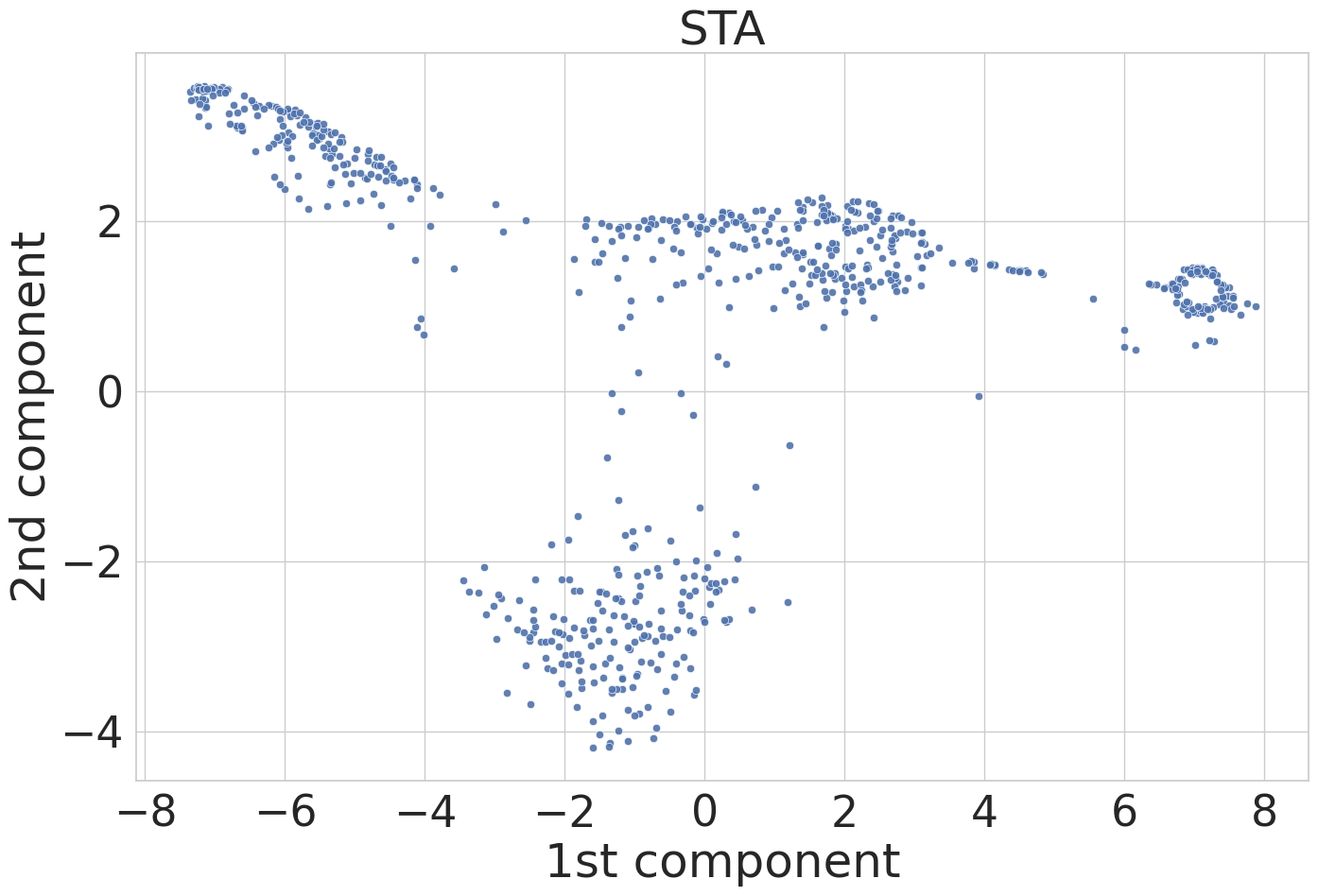}}
  \subfloat[Eve  \label{subfig:eve_mob}]{%
        \includegraphics[ height=3.3cm,width=0.33\columnwidth]{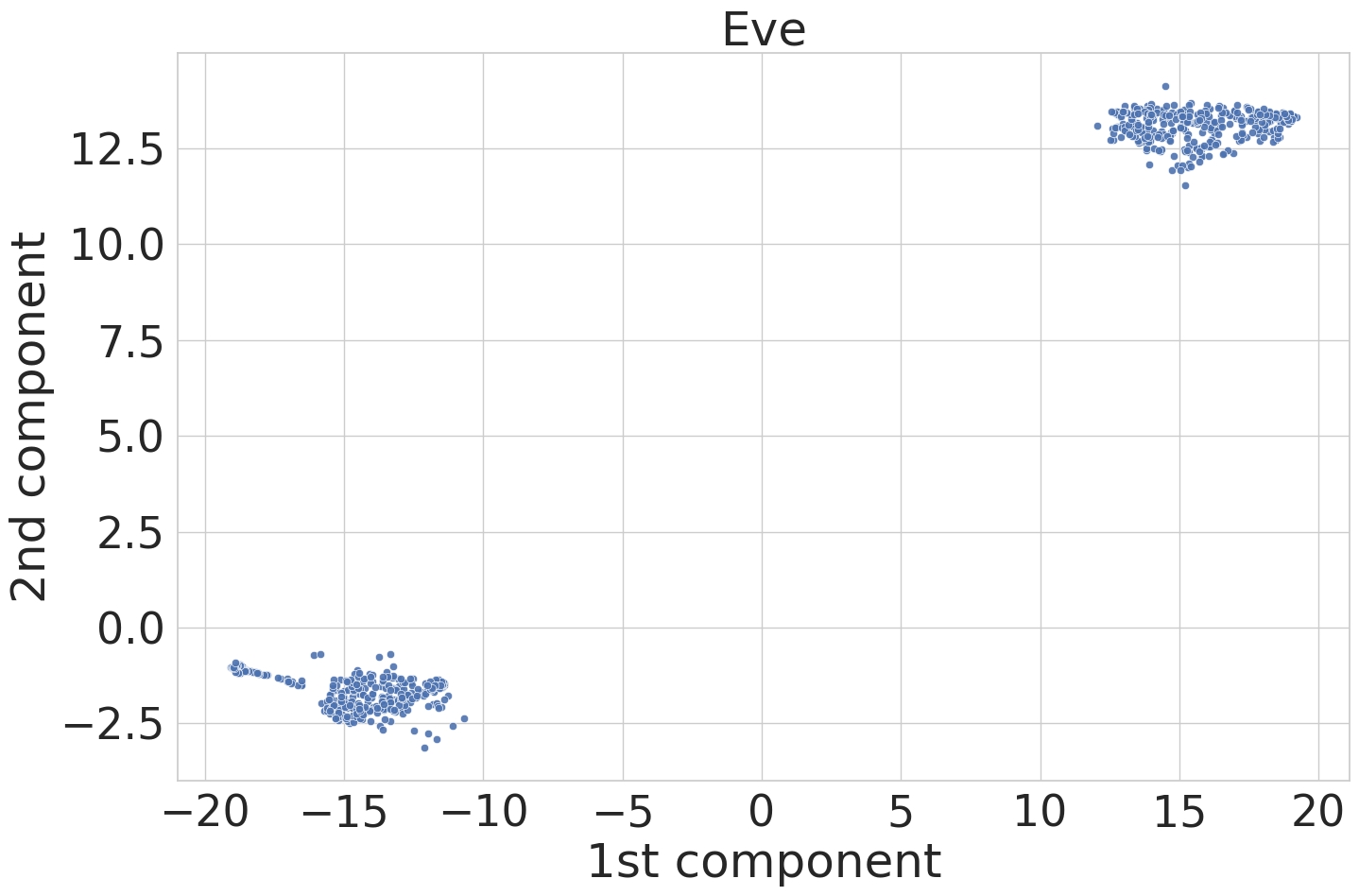}
    }}
   \caption{Clusters structure observed under AP, STA, and Eve under the \mob~for subcarrier $40$ at two timestamps T1 and T2.}
  \label{fig:mob_tsne} 
\end{figure}

Fig.~\ref{KGR} also depicts the impact of the mobility on KGRs. The figure demonstrates that with mobility, the proposed HMM-aided technique generates more keys with bit-mismatch values $\geq 20 \%$ compared to the static scenario. 
The increase in KGR under \mob~is attributed to more clusters in the CSI scattering features embeddings. By inspecting the clusters sets observed by AP and STA under the \mob, we observed more clusters at AP and STA. For instance, in Fig.~\ref{fig:mob_tsne} which shows the CSI scattering embeddings for subcarrier $40$ at two different timestamps under \mob, we observe $4$ clusters at AP and STA compared to $3$ under \static in Fig.~\ref{fig:tsne2}. 
Table~\ref{Tb1} shows the average BER for each studied technique for the \static~and \mob. The table shows comparable BERs achieved under the proposed and AE techniques, which is about $0.2$, compared to a higher BER of  $0.27$ under the BCFL technique for the \static. For the \mob, the table shows lower BERs for the proposed technique compared to AE and BCFL. The table also shows an overall expected degraded BER under the \mob~compared to under the \static~for all techniques highlighting the negative impact of channel dynamics on BER.

%
%

\begin{table} 
\centering 
\caption{ Average BER for AP-STA keys.}
\resizebox{1\columnwidth}{!}{\begin{tabular}{|l|c|c|c|} 
       \hline 
       Technique  & \textbf{Proposed} & \textbf{Denoising AE} & \textbf{BCFL}\\
       \hline
       \textbf{Static}  & 0.19 & 0.2 & 0.27 \\
       \hline
       \textbf{Mobile}  & 0.26 & 0.38 & 0.3 \\
       \hline
\end{tabular}}
\label{Tb1}
\end{table}

%

\begin{table*}
\centering 
\caption{Key randomness for different encoding scheme.${^\dag}$ represents permuted keys.}
\resizebox{1\textwidth}{!}{ \begin{tabular}{|l|c|c|c|c|c|c|c|c|} 
       \hline 
       Technique  & One-hot & One-hot${^\dag}$ & Huffman & Huffman${^\dag}$ & Gray & Gray${^\dag}$ & 2-level Gray & 2-level Gray${^\dag}$ \\
       \hline
       Entropy & Fail & Fail  & Fail & Pass  & Fail & Pass & Fail & Pass \\
       \hline
       Frequency & Fail & Fail & Pass &  Pass & Pass& Pass & Pass & Pass\\
       \hline 
       Block Frequency& Fail & Fail & Fail & Pass & Fail& Pass &Fail & Pass \\
       \hline
       Cumulative sum & Fail & Fail & Pass & Pass & Pass & Pass  &Pass& Pass \\
       \hline
       Runs & Fail & Fail & Fail & Pass & Fail &Pass & Fail& Pass \\
       \hline
       Runs of ones & Fail  & Fail & Pass & Pass & Fail& Pass& Fail &Pass \\
       \hline
       FFT & Fail & Fail & Fail & Fail & Pass & Fail& Fail&Pass \\
       \hline
       Serial & Fail & Fail & Fail & Fail & Fail & Fail& Fail &Pass \\
       \hline
\end{tabular}}
\label{Tb2}
\end{table*}

\subsection{Encoding \& Key Randomness Analysis}
Our evaluation shows that the randomness of the keys generated at AP and STA, as assessed by the NIST randomness test \cite{8966}, is highly dependent on the encoding technique. Table~\ref{Tb2} shows the results for nine tests of the NIST test suite obtained using $3$ encoding schemes, One-hot encoding~\cite{bishop2006chapter9}, Huffman encoding~\cite{huffman_method_1952}, and Gray encoding~\cite{bitner_efficient_1976}, for the keys and the permuted keys. For each test, we set the level of significance to $0.01$, thus sequences passing the tests are random with a confidence of $99.9\%$. The results show that one-hot encoded sequences fail the randomness tests even when the sequences are permuted. This is due to the sparsity of one-hot encoded sequences and the presence of multiple sequences of zeros whose length increases with the number of states. The results also demonstrate that Gray encoding produces sequences that pass most of the tests compared to Huffman encoding. Motivated by the performance of Gray encoding, we propose a switching Gray encoding technique to improve the keys randomness.

The intuition behind the switching encoding technique is to enhance key randomness and eliminate periodic patterns by assigning two distinct Gray codes to each state and alternating between them. This ensures that consecutive occurrences of the same state in the key are encoded differently, reducing repetition and preventing predictable sequences.

\section{Conclusion}
\label{conc}
We propose a novel key generation technique leveraging a new perspective of channel reciprocity that shifts from relying on instantaneous channel measurements to leveraging statistical channel reciprocity. By working on the low-dimensional embeddings of the CSI scattering feature representations at end devices, we presented a hidden Markov model-aided key generation and agreement technique that achieves a 5$\times$ higher key generation rate with no bit mismatch compared to traditional benchmarks. Additionally, we propose a new CSI feature extraction method using wavelet scattering networks, which demonstrates resilience to noise and time warping, offering a robust foundation for secret key generation.

\section{Acknowledgment} 
This work is supported in part by NSF Award No. 2350214.

\bibliographystyle{IEEEtran}
\bibliography{IEEEabrv,References}

\end{document}